\documentclass[letterpaper,twocolumn,prl,aps,superscriptaddress,floatfix]{revtex4}
\usepackage{mathptmx}

\usepackage[latin9]{inputenc}
\usepackage{amsmath}
\usepackage{amssymb}
\usepackage{graphicx}
\usepackage{esint}
\usepackage{subfigure}

\makeatletter
\pdfpageheight\paperheight
\pdfpagewidth\paperwidth

\@ifundefined{textcolor}{}
{%
 \definecolor{BLACK}{gray}{0}
 \definecolor{WHITE}{gray}{1}
 \definecolor{RED}{rgb}{1,0,0}
 \definecolor{GREEN}{rgb}{0,1,0}
 \definecolor{BLUE}{rgb}{0,0,1}
 \definecolor{CYAN}{cmyk}{1,0,0,0}
 \definecolor{MAGENTA}{cmyk}{0,1,0,0}
 \definecolor{YELLOW}{cmyk}{0,0,1,0}
}

\usepackage{xcolor}\usepackage{soul}

\newcommand{\bra}[1]{\ensuremath{\left\langle#1\right|}}
\newcommand{\ket}[1]{\ensuremath{\left|#1\right\rangle}}
\definecolor{blue}{rgb}{0,0,1}
\definecolor{red}{rgb}{1,0,0}
\definecolor{green}{rgb}{0,1,0}

\makeatother

\begin{document}

\title{Single-loop realization of arbitrary non-adiabatic holonomic single-qubit quantum gates in a superconducting circuit}

\author{Y. Xu}
\author{W. Cai}
\author{Y. Ma}
\author{X. Mu}
\author{L. Hu}
\affiliation{Center for Quantum Information, Institute for Interdisciplinary Information
Sciences, Tsinghua University, Beijing 100084, China}

\author{Tao Chen}
\affiliation{Guangdong Provincial Key Laboratory of Quantum Engineering and Quantum Materials, and School of Physics\\ and Telecommunication Engineering, South China Normal University, Guangzhou 510006, China}

\author{H. Wang}
\author{Y.P. Song}
\affiliation{Center for Quantum Information, Institute for Interdisciplinary Information
Sciences, Tsinghua University, Beijing 100084, China}

\author{Zheng-Yuan Xue} \email{zyxue83@163.com}
\affiliation{Guangdong Provincial Key Laboratory of Quantum Engineering and Quantum Materials, and School of Physics\\ and Telecommunication Engineering, South China Normal University, Guangzhou 510006, China}

\author{Zhang-qi Yin}\email{yinzhangqi@tsinghua.edu.cn}

\author{L.~Sun}
\email{luyansun@tsinghua.edu.cn}
\affiliation{Center for Quantum Information, Institute for Interdisciplinary Information
Sciences, Tsinghua University, Beijing 100084, China}

%\date{\today}
\begin{abstract}
Geometric phases are noise-resilient, and thus provide a robust way towards high-fidelity quantum manipulation. Here we experimentally demonstrate arbitrary non-adiabatic holonomic single-qubit quantum gates for both a superconducting transmon qubit and a microwave cavity in a single-loop way. In both cases, an auxiliary state is utilized, and two resonant microwave drives are simultaneously applied with well-controlled but varying amplitudes and phases for the arbitrariness of the gate. The resulting gates on the transmon qubit achieve a fidelity of 0.996 characterized by randomized benchmarking and the ones on the cavity show an averaged fidelity of 0.978 based on a full quantum process tomography. In principle, a nontrivial two-qubit holonomic gate between the qubit and the cavity can also be realized based on our presented experimental scheme. Our experiment thus paves the way towards practical non-adiabatic holonomic quantum manipulation with both qubits and cavities in a superconducting circuit.
\end{abstract}
\maketitle

\vskip 0.5cm

\narrowtext

High-fidelity quantum manipulation is essential for large scale quantum computation. However, as quantum systems are fragile under noises from either the surrounding environment or the control fields, error-resilient manipulations of quantum states are preferable. Geometric phases~\cite{berry1984quantal,wilczek1984appearance} depend only on the global properties of the evolution trajectories, and thus have built-in noise-resilient features against certain local noises~\cite{zhu2005geometric,solinas2012on,johansson2012robustness,Berger2013Exploring,Yale2016Optical}. Therefore, they can naturally be used to achieve high-fidelity quantum gates. Consequently, considerable interests have been paid to various applications of geometric phases in quantum computation~\cite{sjoqvist2008trend}.

Due to the non-commutativity, non-Abelian geometric phases are natural for the so-called holonomic quantum computation ~\cite{zanardi1999holonomic}. Schemes based on the adiabatic evolution of the non-Abelian geometric phases have been proposed on a variety of systems for quantum computation~\cite{duan2001geometric,pachos1999non,wu2005holonomic,Yin2007Implementation,kamleitner2011geometric,albert2016holonomic}. However, these schemes are rather difficult for experimental realization as they rely on complicated control over multilevel systems. Meanwhile, the gates are based on the rather slow adiabatic quantum dynamics and thus decoherence can induce considerable errors. Therefore, it is desirable to implement quantum gates with non-adiabatic evolutions~\cite{aharonov1987phase}. Recently, much attention has been paid to the non-adiabatic holonomic quantum computation with three-level systems~\cite{sjoqvist2012non,xu2012nonadiabatic}. Compared to the adiabatic ones, this type of new schemes is fast and easy to realize, and has been experimentally demonstrated in superconducting circuits~\cite{abdumalikov2013experimental,Egger2018Entanglement}, NMR~\cite{Feng2013Experimental}, and  nitrogen-vacancy centers in diamond~\cite{zu2014experimental,arroyocamejo2014room,yuhei2017optical}.

More importantly, arbitrary single-qubit holonomic gates can be achieved in a single-loop evolution, i.e., a single closed loop evolution in parameter space~\cite{xu2015nonadiabatic,herterich2016single,hong2017implementing}. This will simplify gate sequences in practical quantum information processing compared with the original proposal~\cite{sjoqvist2012non}, where two sequential gates are required for an arbitrary single-qubit gate. In the last year, the single-loop scheme has been experimentally demonstrated in NMR~\cite{li2017experimental} and nitrogen-vacancy centers in diamond with off-resonance drives~\cite{zhou2017holonomic}, which are basically incompatible with pulse shaping and experimentally difficult. Therefore, single-loop schemes with resonant drives are generally required to use shaped pulses to reduce errors.

Here, with a circuit quantum electrodynamics architecture~\cite{Wallraff,Clarke2008Superconducting,You2011Atomic,devoret2013superconducting,Gu2017Microwave} we experimentally demonstrate arbitrary non-adiabatic holonomic single-qubit gates for both a superconducting transmon qubit and a microwave photonic qubit in a single-loop way. This is realized by varying the amplitudes and phases of a two-tone resonant microwave drive~\cite{hong2017implementing}. Besides transmon qubits, photonic qubits in a microwave cavity are also desirable for quantum information processing because of their long coherence times~\cite{Reagor,Reagor2016} and ease of realizing quantum error correction~\cite{LeghtasPRL2013,Michael2016}. In our realization, the gates on the transmon qubit achieve a fidelity of 0.996 characterized by randomized benchmarking (RB), also consistent with the results from a full quantum process tomography (QPT); the gates on the cavity show an averaged fidelity of 0.978 based on a full QPT. Besides local noises, the demonstrated holonomic gates on the qubit are also robust against control amplitude errors and qubit frequency shifts induced by crosstalk~\cite{Supplement}, which become prominent as qubit coherence properties are improved and the size of quantum system increases. The holonomic gates on the cavity provide an alternative way of arbitrary control over Fock states, which also could be robust against experimental noises as the ones on the qubit.

\begin{figure}
\includegraphics{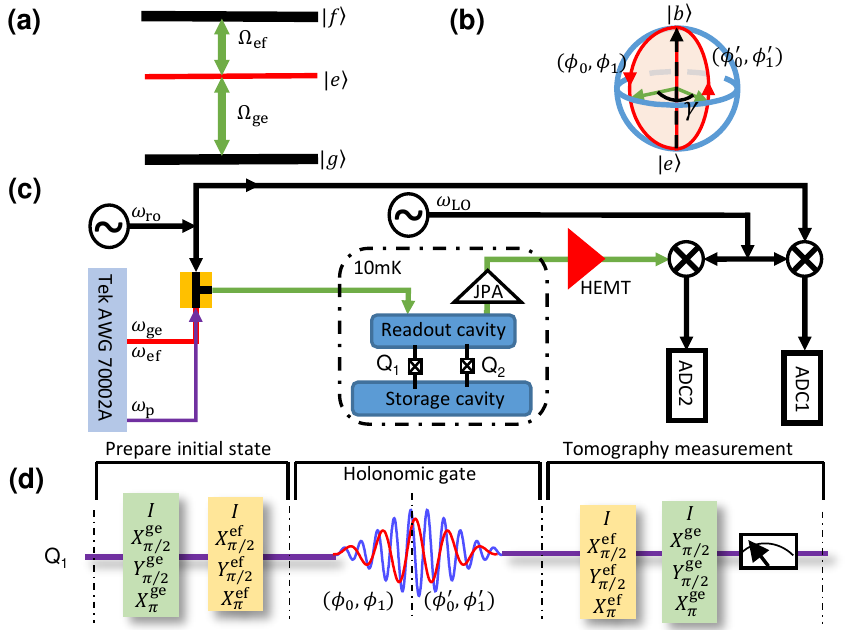}
\caption{Single-loop single-qubit holonomic gates. (a) Two microwave fields are resonantly coupled to the $\ket g \leftrightarrow \ket e$ and $\ket e \leftrightarrow \ket{f}$ transitions of a transmon qubit to generate arbitrary single-qubit holonomic gates in the \{$\ket{g}$ , $\ket{f}$\} subspace. The first excited state $\ket{e}$ is an auxiliary level. (b) Bloch sphere representation of the holonomic gate: a combination of two microwave fields, with two different sets of phases for the first and the second half of the gate operation, equivalently drives the bright state $\ket b$ to the auxiliary state $\ket e$ and then back with an additional phase, while the dark state $\ket d$ remains unchanged. (c) Simplified experimental setup. Two transmon qubits in two microwave trenches are coupled to two microwave cavities, one for readout and the other one for storage. The qubit holonomic gates are demonstrated with $Q_1$, while the cavity holonomic gates are realized with the Fock states \{$\ket{0}$, $\ket{1}$\} in the storage cavity facilitated by both qubits (Fig.~\ref{fig:fig4}). (d) QPT is used to characterize the performance of the arbitrary holonomic gates. Arbitrary initial states are prepared with sequential pulses on $\ket g \leftrightarrow \ket e$ and $\ket e \leftrightarrow \ket f$ transitions. Nine sequential pre-rotation pulses on the qubit are performed before the final measurement to obtain the transmon's full state tomography with three levels.}
\label{fig:fig1}
\end{figure}

We first address the implementation of arbitrary single-qubit holonomic gates on a superconducting transmon qubit in the \{$\ket{g}$ , $\ket{f}$\} subspace, as shown in Fig.~\ref{fig:fig1}a. Here, $\ket g$, $\ket e$, and $\ket f$ denote the three lowest energy levels of the transmon qubit; $\ket e$ is an auxiliary state and remains unoccupied before and after the gate operation. Our scheme consists of two microwave fields resonantly coupled to the sequential transitions $|g\rangle\leftrightarrow|e\rangle$ and $|e\rangle\leftrightarrow|f\rangle$ of the transmon qubit, as described by
\begin{eqnarray}
\label{eq:h1}
\mathcal{H}_{1} &=& \Omega_{\mathrm{ge}}(t)e^{i\phi_0}|g\rangle\langle e|+
\Omega_{\mathrm{ef}}(t)e^{i\phi_1}|f\rangle\langle e|  + \mathrm{H.c.}\notag\\
&=&\Omega (t) e^{i(\phi_1-\pi)} \left(\sin\frac{\theta}{2}e^{i\phi}|g\rangle -\cos\frac{\theta}{2}|f\rangle\right)\langle e| + \mathrm{H.c.},
\end{eqnarray}
where $\Omega_{\mathrm{ge}}(t)$ and $\Omega_{\mathrm{ef}}(t)$ are the time-dependent amplitudes of the two microwave drives with the corresponding initial phases $\phi_0$ and $\phi_1$; $\phi=\phi_0-\phi_1+\pi$, $\Omega(t)=\sqrt{\Omega^2_{\mathrm{ge}}(t)+\Omega^2_{\mathrm{ef}}(t)}$, and $\tan(\theta/2)=\Omega_{\mathrm{ge}}(t)/\Omega_{\mathrm{ef}}(t)$. As seen in Eq.~\ref{eq:h1}, the quantum dynamics is captured by the resonant coupling between the bright state $|b\rangle=\sin(\theta/2)e^{i\phi}|g\rangle-\cos(\theta/2)|f\rangle$ and the auxiliary state $|e\rangle$, while the dark state $|d\rangle=\cos(\theta/2)|g\rangle+\sin(\theta/2)e^{-i\phi}|f\rangle$ is decoupled. Under the cyclic evolution condition, $\int_0^T \Omega(t) dt=\pi$, one can obtain a quantum gate depending on $\theta$ and/or $\phi$. Meanwhile, since there is no transition between $|d\rangle$ and $|b\rangle$ states when $\theta$ is time-independent and also no dynamical phases due to the on-resonance drives, the obtained gates are thus holonomic~\cite{sjoqvist2012non}.

To achieve a universal set of single-qubit holonomic gates in a single-loop way \cite{hong2017implementing}, we divide the evolution time $T$ into two equal halves and choose $\phi_0=\phi$, $\phi_1= \pi$ for $t\in [0, T/2]$ and $\phi_0'=\phi+ \gamma- \pi$, $\phi_1'= \gamma$ for $t\in [T/2, T]$, such that the Hamiltonians during these two halves are $\mathcal{H}_\mathrm{A}=\Omega(t)(|b\rangle\langle e|+ |e\rangle\langle b|)$ and $\mathcal{H}_\mathrm{B}=-\Omega(t) (e^{i\gamma} |b\rangle\langle e|+ e^{-i\gamma} |e\rangle\langle b|)$, respectively. Geometrically, the two evolutions coincide at two poles in the Bloch sphere, and the cyclic geometric phase is illustrated as the red slice contour in Fig.~\ref{fig:fig1}b. Therefore, in the qubit subspace $\{|g\rangle, |f\rangle\}$, the obtained holonomic single-qubit gate is
\begin{eqnarray}\label{u1}
U_1(\theta,\gamma,\phi)&=&\left(\begin{array}{cc}
\cos\frac{\gamma}{2} - i\sin\frac{\gamma}{2}\cos\theta & -i\sin\frac{\gamma}{2}\sin\theta e^{i\phi} \\ -i\sin\frac{\gamma}{2}\sin\theta e^{-i\phi} & \cos\frac{\gamma}{2} + i\sin\frac{\gamma}{2}\cos\theta
                        \end{array}\right)\notag\\
&=& \exp\left({-i{\gamma \over 2} {\bf n}\cdot{\bf \sigma}}\right),
\end{eqnarray}
which describes a rotation operation around the axis ${\bf n}= (\sin\theta\cos\phi, \sin\theta\sin\phi, \cos\theta)$ by an angle $\gamma$, up to a global phase factor $\exp(i\gamma/2)$.

In our experiment, two superconducting transmon qubits are dispersively coupled to two three-dimensional cavities~\cite{Paik,Vlastakis,SunNature,Liu2017}, as shown in Fig.~\ref{fig:fig1}c. The $\ket g \leftrightarrow \ket e$ and $\ket e \leftrightarrow \ket f$ transition frequencies of the two qubits $Q_1$ and $Q_2$ are $\omega_{\mathrm{ge1}}/2\pi=5.036$~GHz and $\omega_{\mathrm{ef1}}/2\pi=4.782$~GHz, $\omega_{\mathrm{ge2}}/2\pi=5.605$~GHz and $\omega_{\mathrm{ef2}}/2\pi=5.367$~GHz, respectively. One of the cavities with a transition frequency of $\omega_{\mathrm{r}}/2\pi=8.540$~GHz is connected to a Josephson parametric amplifier (JPA) for a fast and high-fidelity joint readout of the two-qubit states~\cite{Hatridge,Roy,Kamal,Murch}. The other cavity with a transition frequency of $\omega_{\mathrm{s}}/2\pi=7.614$~GHz is utilized for storage and manipulation of the photonic states and for implementing the holonomic gates between Fock states $\ket 0$ and $\ket 1$ as will be discussed below. In the following, we ignore the readout cavity and the ``cavity" refers to the storage cavity.  More details about the device parameters can be found in Ref.~\cite{Supplement}.

\begin{figure}
\includegraphics{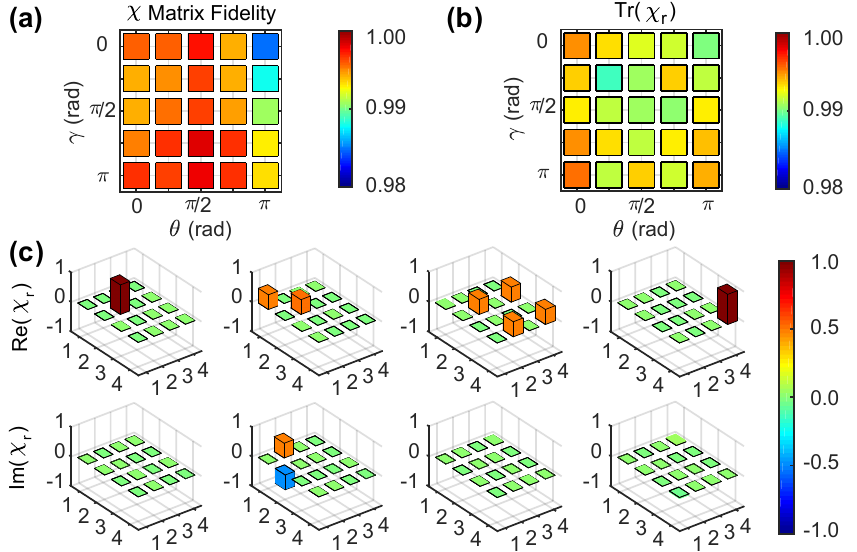}
\caption{QPT of the single-qubit holonomic gates. (a) Unattenuated $\chi$ matrix fidelity $F_{\mathrm{unatt}}$ of the single-qubit holonomic gates $U_1(\theta,\gamma,\phi)$ with different $\theta$ and $\gamma$ while $\phi=0$. The averaged fidelity $\bar{F}_{\mathrm{unatt}}=0.994$ while the averaged attenuated fidelity $\bar{F}_{\mathrm{att}}=0.975$. (b) The traces of the reduced process matrix $\chi_\mathrm{r}$ as a function of both $\theta$ and $\gamma$ with $\phi=0$. The averaged trace is 0.992, indicating small leakage outside the computation subspace \{$\ket g$, $\ket f$\}. (c) Bar charts of the real and imaginary parts of $\chi_\mathrm{r}$ of four specific gates: $X_{\pi}=U_1(\pi/2,\pi,0)$, $X_{\pi/2}=U_1(\pi/2,\pi/2,0)$, $H=U_1(\pi/4,\pi,0)$, and $Z_{\pi}=U_1(0,\pi,0)$, where $R_{\varphi}$ denotes a rotation of the qubit by an angle $\varphi$ along the axis $R$ and $H$ represents the Hadamard gate. The numbers in the $x$ and $y$ axes correspond to the operators in the basis set \{$I$, $X$, $-iY$, $Z$\} of the \{$\ket g$, $\ket f$\} subspace. The solid black outlines are for the ideal gates.}
\label{fig:fig2}
\end{figure}

We now demonstrate the realization of the arbitrary holonomic gates in a single-loop way with transmon qubit $Q_1$ based on the procedure discussed above. The envelopes of the two drives are truncated Gaussian pulses with a total width of $4\sigma=120$~ns. We characterize the holonomic single-qubit gates by a full QPT including all three levels, $\ket{g}$, $\ket{e}$, and $\ket{f}$~\cite{Supplement,thew2002qudit,bianchetti2010control}. The experimental pulse sequence is shown in Fig.~\ref{fig:fig1}d. To evaluate the QPT, we have used both attenuated and unattenuated $\chi$ matrix fidelities, which are respectively defined as $F_{\mathrm{att}}=\left| \texttt{Tr}\left(\chi_{\mathrm{exp}}\chi_{\mathrm{th}}^{\dagger} \right) \right|$ and $F_{\mathrm{unatt}} = {\left| \texttt{Tr}\left( \chi_{\mathrm{exp}}\chi_{\mathrm{th}}^{\dagger} \right)\right|}/{\sqrt{\texttt{Tr}\left( \chi_{\mathrm{exp}}\chi_{\mathrm{exp}}^{\dagger} \right) \texttt{Tr}\left( \chi_{\mathrm{th}}\chi_{\mathrm{th}}^{\dagger} \right)}}$~\cite{Weinstein2004,Zhang2012PRL,Feng2013Experimental}, where $\chi_{\mathrm{exp}}$ is the experimental process matrix and $\chi_{\mathrm{th}}$ is the corresponding ideal process matrix. The latter fidelity can ignore the errors due to signal loss, e.g., the errors in state preparations and measurements. Figure~\ref{fig:fig2}a shows $F_{\mathrm{unatt}}$ of the gates as a function of both $\theta$ and $\gamma$ with $\phi=0$, and the averaged fidelity $\bar{F}_{\mathrm{unatt}}=0.994$. Energy relaxation and dephasing of both excited states and non-perfect microwave drives can cause a population leakage outside the computation subspace \{$\ket g, \ket f$\} to the auxiliary $\ket{e}$ state. This leakage can be characterized by the trace of the reduced process matrix $\chi_\mathrm{r}$, which describes the process only involving $\ket g$ and $\ket f$ and ignores any operators acting on the auxiliary state $\ket{e}$~\cite{Supplement}. Figure~\ref{fig:fig2}b shows the traces of $\chi_\mathrm{r}$ originated from the measured $\chi_\mathrm{exp}$ whose fidelities are shown in Fig.~\ref{fig:fig2}a. The high value (0.992) of the averaged trace indicates that there is nearly no leakage outside the computation subspace for the holonomic gates on the transmon qubit. $\chi_\mathrm{r}$ of four example gates are shown in Fig.~\ref{fig:fig2}c with $F_{\mathrm{unatt}}=$ 0.997, 0.996, 0.996, and 0.996 respectively (the corresponding  $F_{\mathrm{att}}=$ 0.976, 0.980, 0.963, and 0.988).

\begin{figure}
\includegraphics{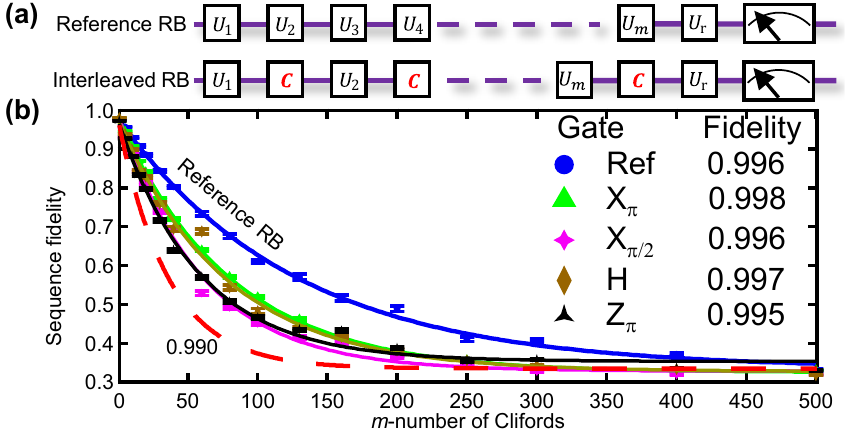}
\caption{RB of the single-qubit holonomic gates. (a) Sequences of both a reference RB experiment and an interleaved RB experiment. (b) The sequence fidelity decay as a function of the gate length $m$. The fidelity for each sequence length $m$ is measured for $k=100$ different random sequences with the standard deviation from the mean plotted as the error bars. Both curves are fitted with $F=Ap^{m}+B$ with different sequence decays $p=p_{\mathrm{ref}}$ and $p_{\mathrm{gate}}$. The average gate fidelity per Clifford gate is $F_{\mathrm{avg}}=1-(1-p_{\mathrm{ref}})/2=0.996$. The difference between the reference and the interleaved RB experiments gives the specific gate fidelity $F_{\mathrm{gate}}=1-(1-p_{\mathrm{gate}}/p_{\mathrm{ref}})/2$. The red dashed line indicates the threshold for exceeding gate fidelity of 0.990.}
\label{fig:fig3}
\end{figure}

Another regular way to extract gate fidelity only without relying on perfect state preparations and measurements is RB~\cite{knill2008randomized,chow2009randomized,magesan2012efficient,magesan2011scalable,barends2014superconducting}. An agreement between $F_{\mathrm{unatt}}$ and the fidelity from RB should provide more confidence on the gate performance. We utilize the Clifford-based RB and the experimental sequences are shown in Fig.~\ref{fig:fig3}a, where we perform both a reference RB experiment and an interleaved RB experiment. The results of the four holonomic gates presented in Fig.~\ref{fig:fig2}c are shown in Fig.~\ref{fig:fig3}b. Each Clifford gate is realized by choosing specific parameters $\theta$, $\gamma$, and $\phi$. The reference RB experiment gives an average gate fidelity of the single-qubit holonomic gates in the Clifford group $F_{\mathrm{avg}}=0.996$. The measured gate fidelities of the four specific holonomic gates $X_{\pi}$, $X_{\pi/2}$, $H$, and $Z_{\pi}$ are 0.998, 0.996, 0.997, and 0.995, respectively. These fidelities are consistent with the measured ${F}_{\mathrm{unatt}}$, thus validating ${F}_{\mathrm{unatt}}$ as a good measure of gate performance. The loss of fidelity is mainly from the decoherence of both $\ket{e}$ and $\ket{f}$ of the transmon qubit, as confirmed by numerical simulations based on QuTiP in Python~\cite{Johansson2012,Johansson2013}.

\begin{figure}
\includegraphics{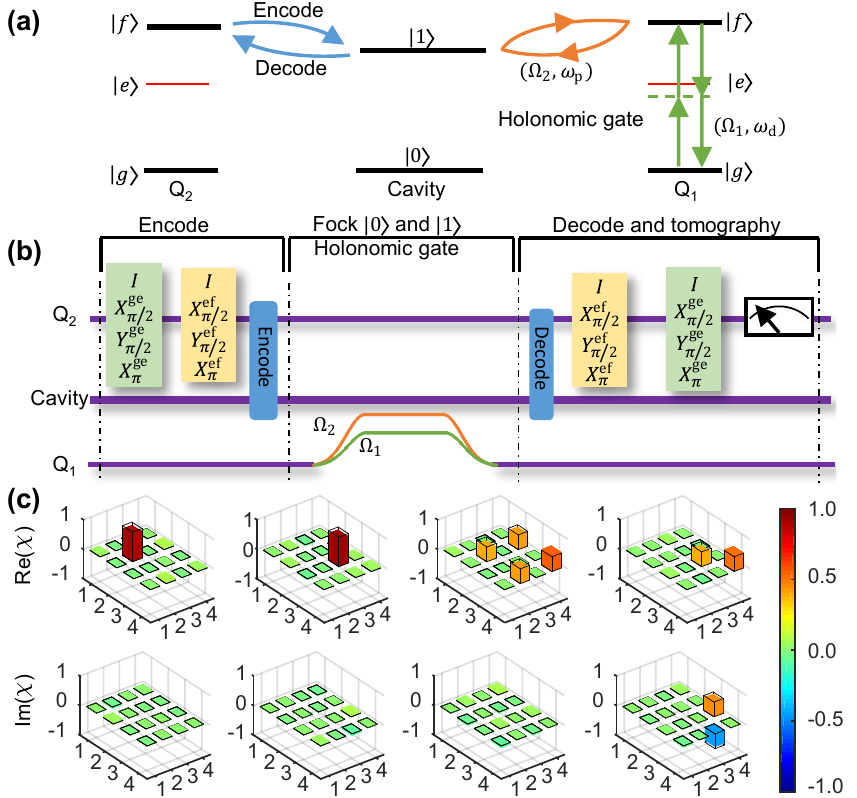}
\caption{Holonomic gates for cavity Fock state subspace \{$\ket 0$, $\ket 1$\}. (a) Illustration of the holonomic gates for Fock states, as well as the encoding and decoding processes. The holonomic gates are implemented by using a selective two-photon transition drive $\Omega_1e^{i\omega_\mathrm{d}t}$ on qubit $Q_1$ and a cavity-assisted Raman transition drive $\Omega_2e^{i\omega_\mathrm{p}t}$ between $\ket{1g}$ and $\ket{0f}$. The encoding and decoding processes are realized with a swap operation between qubit $Q_2$ and the cavity mode through a Raman transition drive similar to that for the gate. (b) Experimental sequence to perform QPT of the holonomic gates in the Fock \{$\ket 0$,$\ket 1$\} subspace. (c) Bar charts of the real and imaginary parts of the process $\chi$ matrices for the whole process, including both the encoding and decoding processes, with the following gates: $X_{\pi}=U_2(\pi/2,0)$, $Y_{\pi}=U_2(\pi/2,\pi/2)$, $H_1=U_2(\pi/4,0)$, and $H_2=U_2(\pi/4,\pi/2)$. The attenuated and unattenuated process fidelities for these gates are: 0.880, 0.881, 0.877, 0.879 and 0.990, 0.990, 0.970, 0.962, respectively. The numbers in the $x$ and $y$ axes correspond to the operators in the basis set \{$I$, $X$, $-iY$, $Z$\}. The solid black outlines are for the ideal gates.}
\label{fig:fig4}
\end{figure}

In addition to the implementation of holonomic gates on the transmon qubit, we also realize holonomic operations on the cavity Fock states following a similar scheme. As shown in Fig.~\ref{fig:fig4}a, the holonomic gates are implemented by using a selective two-photon transition drive $\Omega_1e^{i\omega_\mathrm{d}t}$ on qubit $Q_1$ conditional on only zero photon in the cavity and a cavity-assisted Raman transition drive $\Omega_2e^{i\omega_\mathrm{p}t}$ between $\ket{1g}$ and $\ket{0f}$~\cite{zeytinoglu2015microwave,pechal2014microwave}. Here the drive frequencies $\omega_\mathrm{d}=\left(\omega_{\mathrm{ge1}}+\omega_{\mathrm{ef1}}\right)/2$, $\omega_\mathrm{p}=\omega_{\mathrm{ge1}}+\omega_{\mathrm{ef1}}-\omega_\mathrm{s}$, and $\Omega_1$ and $\Omega_2$ are the corresponding drive strengths. In the joint state notation, the numbers represent the Fock states in the cavity and the letters correspond to the states of qubit $Q_1$. 

Similarly, the above two drives together generate the following effective Hamiltonian~\cite{hong2017implementing,zeytinoglu2015microwave}:
\begin{eqnarray} \label{h2}
\mathcal{H}_{2} &=& \tilde{g}_1\ket{0g}\bra{0f}+ \tilde{g}_2\ket{1g}\bra{0f}  + \mathrm{H.c.}\notag\\
&=& \tilde{g} \left(\sin\frac{\theta}{2}e^{i\phi}\ket{0g} -\cos\frac{\theta}{2}\ket{1g}\right)\bra{0f} + \mathrm{H.c.},
\end{eqnarray}
where, $\tilde{g}_1$ and $\tilde{g}_2$ are the effective coupling strengthes of the two drives, and $\tilde{g}=\sqrt{\tilde{g}_1^2+\tilde{g}_2^2}$, ${\tilde{g}_1}/{\tilde{g}_2}=\tan(\theta/2)e^{i\phi}$. To validate the above Hamiltonian, $\tilde{g}_1$ of the two-photon transition drive should be much smaller than the dispersive shifts between qubit $Q_1$ and the cavity. In the experiment, both the two-photon transition drive and the Raman transition drive are set to have the same pulse envelope (a wide square pulse with sine squared ramp-up and ramp-down edges) to keep the two drives synchronized and the ratio of two amplitudes fixed. Both $\tilde{g}_1$ and $\tilde{g}_2$ are carefully calibrated with square pulses with different amplitudes~\cite{Supplement}. We have also carefully taken into account the AC-Stark shifts under the strong external drives to eliminate the possible dynamical phases~\cite{Supplement}. Similarly to the transmon qubit holonomic gates, by adjusting the ratios of $\tilde{g}_1$ and $\tilde{g}_2$, which lead to different $\theta$ and $\phi$, we can realize arbitrary holonomic gates $U_1(\theta,\gamma,\phi)$ on the basis \{$\ket{0g}$, $\ket{1g}$\}, i.e., the Fock state basis \{$\ket{0}$, $\ket{1}$\}. As a demonstration, here we fix $\gamma=\pi$ and realize the holonomic gate
$
U_2(\theta,\phi)=\left(\begin{array}{cc}
\cos\theta & \sin\theta e^{i\phi} \\ \sin\theta e^{-i\phi} & -\cos\theta
                        \end{array}\right).
$

We characterize $U_2(\theta,\phi)$ with full QPT as well, and the results are shown in Fig.~\ref{fig:fig4}b. To make the preparation and characterization of the cavity state easier, we instead use qubit $Q_2$ to facilitate an encoding and decoding process of the cavity state. The encoding and decoding processes are realized with a similar Raman transition drive between qubit $Q_2$ and the cavity as illustrated in Fig.~\ref{fig:fig4}a. We choose four different gates $X_{\pi}$, $Y_{\pi}$, $H_1$, and $H_2$, and perform QPT of these gates. The experimental process matrices $\chi_{\mathrm{exp}}$ are shown in Fig.~\ref{fig:fig4}c. The averaged experimental attenuated and unattenuated process fidelities of these holonomic gates are $\bar{F}_{\mathrm{att}}=0.879$ and $\bar{F}_{\mathrm{unatt}}=0.978$, respectively. The infidelities of the holonomic gates for Fock states are mainly limited by the encoding and decoding errors (including the initial state preparation and final measurement errors), decoherence process during the gate, and imperfections of the control pulses. As a reference, the process fidelity of the encoding and decoding processes only without any gate is 0.95 (an error of 5\%). The dissipation and dephasing of the two excited states $\ket{e}$ and $\ket{f}$ during the gate process can induce an additional infidelity of 5-6\% due to the long gate duration time. These two dominant error budgets are consistent with the measured attenuated fidelities. A higher fidelity gate can be achieved with a shorter gate operation time under a larger dispersive shift.

To achieve a universal quantum computation, two-qubit gates are necessary. A nontrivial two-qubit holonomic gate between the qubit and the cavity can in principle be realized in a similar way to the single-qubit gates, i.e., by using two resonant selective pulses on $\ket g \leftrightarrow \ket e$ and $\ket e \leftrightarrow \ket f$ transitions respectively, conditional on only zero photon in the cavity. Then the effective Hamiltonian is $\mathcal{H}_3 = \Omega_{\mathrm{ge}}\left(t\right)e^{i\phi_0}\ket{0g}\bra{0e} + \Omega_{\mathrm{ef}}e^{i\phi_1}\left(t\right)\ket{0f}\bra{0e} + \text{H.c.}$.
A nontrivial two-qubit holonomic gate in the subspace $\{\ket{0g}, \ket{0f}, \ket{1g}, \ket{1f} \}$ can then be realized with the form of $
U_3\left(\theta,\gamma,\phi \right) = \left(  \begin{array}{cc}
 U_1\left(\theta,\gamma,\phi \right) & 0 \\
 0 & I \\
    \end{array}\notag  \right)
$.

In conclusion, in a circuit quantum electrodynamics architecture we have experimentally demonstrated high-fidelity arbitrary non-adiabatic holonomic single-qubit gates for both a superconducting transmon qubit and a microwave cavity in a single-loop way. Moreover, our method can be generalized to achieve holonomic gates between the nearest Fock states $\ket n$ and $\ket{n+1}$, if the resonant drives are on $\ket{n,g}\leftrightarrow\ket{n,f}$ and $\ket{n+1,g}\leftrightarrow\ket{n,f}$ transitions. Combining these gates, we can realize full holonomic control of both the transmon qubit and the cavity mode~\cite{Vogel1993,Law1996,Hofheinz2009,Heeres2015,Wang2017}, which has important applications in cavity-assisted quantum information processing and high-precision measurements~\cite{Caves1980}. Our experiment thus opens the door to implement holonomic manipulations of both qubits and cavities in a superconducting circuit. In addition, the \{$\ket{g}$, $\ket{f}$\} encoding is critical for the recently realized distributed quantum information processing~\cite{pechal2014microwave,Kurpiers2017Deterministic} and fault-tolerant measurement of a quantum error syndrome~\cite{Rosenblum2018Fault}, our holonomic gates therefore can be important and readily used in these two directions.

\begin{acknowledgments}
This work was supported by the National
Key R\&D Program of China (Grants  No.2017YFA0304303 and No. 2016YFA0301803) and National Natural Science
Foundation of China (Grants  No.11474177,  No. 11874156, and No. 61771278). L.S. also acknowledges the support the Thousand Youth Fellowship program in China. L.S. also thanks R. Vijay and his group for help on the parametric amplifier
measurements.
\end{acknowledgments}

\textit{Note added.} -- Recently, we became aware of a similar implementation in a different system~\cite{Ishida2018Universal}, but with a lower average gate fidelity due to the short coherence times of the auxiliary excited state.

%\bibliography{bibliography}

\begin{thebibliography}{69}
\expandafter\ifx\csname natexlab\endcsname\relax\def\natexlab#1{#1}\fi
\expandafter\ifx\csname bibnamefont\endcsname\relax
  \def\bibnamefont#1{#1}\fi
\expandafter\ifx\csname bibfnamefont\endcsname\relax
  \def\bibfnamefont#1{#1}\fi
\expandafter\ifx\csname citenamefont\endcsname\relax
  \def\citenamefont#1{#1}\fi
\expandafter\ifx\csname url\endcsname\relax
  \def\url#1{\texttt{#1}}\fi
\expandafter\ifx\csname urlprefix\endcsname\relax\def\urlprefix{URL }\fi
\providecommand{\bibinfo}[2]{#2}
\providecommand{\eprint}[2][]{\url{#2}}

\bibitem[{\citenamefont{Berry}(1984)}]{berry1984quantal}
\bibinfo{author}{\bibfnamefont{M.~V.} \bibnamefont{Berry}},
  \bibinfo{journal}{Proc. R. Soc. Lond.} \textbf{\bibinfo{volume}{392}},
  \bibinfo{pages}{45} (\bibinfo{year}{1984}).

\bibitem[{\citenamefont{Wilczek and Zee}(1984)}]{wilczek1984appearance}
\bibinfo{author}{\bibfnamefont{F.}~\bibnamefont{Wilczek}} \bibnamefont{and}
  \bibinfo{author}{\bibfnamefont{A.}~\bibnamefont{Zee}},
  \bibinfo{journal}{Phys. Rev. Lett.} \textbf{\bibinfo{volume}{52}},
  \bibinfo{pages}{2111} (\bibinfo{year}{1984}).

\bibitem[{\citenamefont{Zhu and Zanardi}(2005)}]{zhu2005geometric}
\bibinfo{author}{\bibfnamefont{S.-L.} \bibnamefont{Zhu}} \bibnamefont{and}
  \bibinfo{author}{\bibfnamefont{P.}~\bibnamefont{Zanardi}},
  \bibinfo{journal}{Phys. Rev. A} \textbf{\bibinfo{volume}{72}},
  \bibinfo{pages}{020301} (\bibinfo{year}{2005}).

\bibitem[{\citenamefont{Solinas et~al.}(2012)\citenamefont{Solinas, Sassetti,
  Truini, and Zangh\`{\i}}}]{solinas2012on}
\bibinfo{author}{\bibfnamefont{P.}~\bibnamefont{Solinas}},
  \bibinfo{author}{\bibfnamefont{M.}~\bibnamefont{Sassetti}},
  \bibinfo{author}{\bibfnamefont{P.}~\bibnamefont{Truini}}, \bibnamefont{and}
  \bibinfo{author}{\bibfnamefont{N.}~\bibnamefont{Zangh\`{\i}}},
  \bibinfo{journal}{New J. Phys.} \textbf{\bibinfo{volume}{14}},
  \bibinfo{pages}{093006} (\bibinfo{year}{2012}).

\bibitem[{\citenamefont{Johansson
  et~al.}(2012{\natexlab{a}})\citenamefont{Johansson, Sj\"oqvist, Andersson,
  Ericsson, Hessmo, Singh, and Tong}}]{johansson2012robustness}
\bibinfo{author}{\bibfnamefont{M.}~\bibnamefont{Johansson}},
  \bibinfo{author}{\bibfnamefont{E.}~\bibnamefont{Sj\"oqvist}},
  \bibinfo{author}{\bibfnamefont{L.~M.} \bibnamefont{Andersson}},
  \bibinfo{author}{\bibfnamefont{M.}~\bibnamefont{Ericsson}},
  \bibinfo{author}{\bibfnamefont{B.}~\bibnamefont{Hessmo}},
  \bibinfo{author}{\bibfnamefont{K.}~\bibnamefont{Singh}}, \bibnamefont{and}
  \bibinfo{author}{\bibfnamefont{D.~M.} \bibnamefont{Tong}},
  \bibinfo{journal}{Phys. Rev. A} \textbf{\bibinfo{volume}{86}},
  \bibinfo{pages}{062322} (\bibinfo{year}{2012}{\natexlab{a}}).

\bibitem[{\citenamefont{Berger et~al.}(2013)\citenamefont{Berger, Pechal,
  Abdumalikov, Eichler, Steffen, Fedorov, Wallraff, and
  Filipp}}]{Berger2013Exploring}
\bibinfo{author}{\bibfnamefont{S.}~\bibnamefont{Berger}},
  \bibinfo{author}{\bibfnamefont{M.}~\bibnamefont{Pechal}},
  \bibinfo{author}{\bibfnamefont{A.~A.} \bibnamefont{Abdumalikov}},
  \bibinfo{author}{\bibfnamefont{C.}~\bibnamefont{Eichler}},
  \bibinfo{author}{\bibfnamefont{L.}~\bibnamefont{Steffen}},
  \bibinfo{author}{\bibfnamefont{A.}~\bibnamefont{Fedorov}},
  \bibinfo{author}{\bibfnamefont{A.}~\bibnamefont{Wallraff}}, \bibnamefont{and}
  \bibinfo{author}{\bibfnamefont{S.}~\bibnamefont{Filipp}},
  \bibinfo{journal}{Phys. Rev. A} \textbf{\bibinfo{volume}{87}},
  \bibinfo{pages}{060303} (\bibinfo{year}{2013}).

\bibitem[{\citenamefont{Yale et~al.}(2016)\citenamefont{Yale, Heremans, Zhou,
  Auer, Burkard, and Awschalom}}]{Yale2016Optical}
\bibinfo{author}{\bibfnamefont{C.~G.} \bibnamefont{Yale}},
  \bibinfo{author}{\bibfnamefont{F.~J.} \bibnamefont{Heremans}},
  \bibinfo{author}{\bibfnamefont{B.~B.} \bibnamefont{Zhou}},
  \bibinfo{author}{\bibfnamefont{A.}~\bibnamefont{Auer}},
  \bibinfo{author}{\bibfnamefont{G.}~\bibnamefont{Burkard}}, \bibnamefont{and}
  \bibinfo{author}{\bibfnamefont{D.~D.} \bibnamefont{Awschalom}},
  \bibinfo{journal}{Nature Photon.} \textbf{\bibinfo{volume}{10}},
  \bibinfo{pages}{184} (\bibinfo{year}{2016}).

\bibitem[{\citenamefont{Sj\"oqvist}(2008)}]{sjoqvist2008trend}
\bibinfo{author}{\bibfnamefont{E.}~\bibnamefont{Sj\"oqvist}},
  \bibinfo{journal}{Physics} \textbf{\bibinfo{volume}{1}}, \bibinfo{pages}{35}
  (\bibinfo{year}{2008}).

\bibitem[{\citenamefont{Zanardi and Rasetti}(1999)}]{zanardi1999holonomic}
\bibinfo{author}{\bibfnamefont{P.}~\bibnamefont{Zanardi}} \bibnamefont{and}
  \bibinfo{author}{\bibfnamefont{M.}~\bibnamefont{Rasetti}},
  \bibinfo{journal}{Physics Letters A} \textbf{\bibinfo{volume}{264}},
  \bibinfo{pages}{94 } (\bibinfo{year}{1999}), ISSN \bibinfo{issn}{0375-9601}.

\bibitem[{\citenamefont{Duan et~al.}(2001)\citenamefont{Duan, Cirac, and
  Zoller}}]{duan2001geometric}
\bibinfo{author}{\bibfnamefont{L.}~\bibnamefont{Duan}},
  \bibinfo{author}{\bibfnamefont{J.~I.} \bibnamefont{Cirac}}, \bibnamefont{and}
  \bibinfo{author}{\bibfnamefont{P.}~\bibnamefont{Zoller}},
  \bibinfo{journal}{Science} \textbf{\bibinfo{volume}{292}},
  \bibinfo{pages}{1695} (\bibinfo{year}{2001}).

\bibitem[{\citenamefont{Pachos et~al.}(1999)\citenamefont{Pachos, Zanardi, and
  Rasetti}}]{pachos1999non}
\bibinfo{author}{\bibfnamefont{J.}~\bibnamefont{Pachos}},
  \bibinfo{author}{\bibfnamefont{P.}~\bibnamefont{Zanardi}}, \bibnamefont{and}
  \bibinfo{author}{\bibfnamefont{M.}~\bibnamefont{Rasetti}},
  \bibinfo{journal}{Phys. Rev. A} \textbf{\bibinfo{volume}{61}},
  \bibinfo{pages}{010305} (\bibinfo{year}{1999}).

\bibitem[{\citenamefont{Wu et~al.}(2005)\citenamefont{Wu, Zanardi, and
  Lidar}}]{wu2005holonomic}
\bibinfo{author}{\bibfnamefont{L.-A.} \bibnamefont{Wu}},
  \bibinfo{author}{\bibfnamefont{P.}~\bibnamefont{Zanardi}}, \bibnamefont{and}
  \bibinfo{author}{\bibfnamefont{D.~A.} \bibnamefont{Lidar}},
  \bibinfo{journal}{Phys. Rev. Lett.} \textbf{\bibinfo{volume}{95}},
  \bibinfo{pages}{130501} (\bibinfo{year}{2005}).

\bibitem[{\citenamefont{Yin et~al.}(2007)\citenamefont{Yin, Li, and
  Peng}}]{Yin2007Implementation}
\bibinfo{author}{\bibfnamefont{Z.-q.} \bibnamefont{Yin}},
  \bibinfo{author}{\bibfnamefont{F.-l.} \bibnamefont{Li}}, \bibnamefont{and}
  \bibinfo{author}{\bibfnamefont{P.}~\bibnamefont{Peng}},
  \bibinfo{journal}{Phys. Rev. A} \textbf{\bibinfo{volume}{76}},
  \bibinfo{pages}{062311} (\bibinfo{year}{2007}).

\bibitem[{\citenamefont{Kamleitner et~al.}(2011)\citenamefont{Kamleitner,
  Solinas, M\"uller, Shnirman, and M\"ott\"onen}}]{kamleitner2011geometric}
\bibinfo{author}{\bibfnamefont{I.}~\bibnamefont{Kamleitner}},
  \bibinfo{author}{\bibfnamefont{P.}~\bibnamefont{Solinas}},
  \bibinfo{author}{\bibfnamefont{C.}~\bibnamefont{M\"uller}},
  \bibinfo{author}{\bibfnamefont{A.}~\bibnamefont{Shnirman}}, \bibnamefont{and}
  \bibinfo{author}{\bibfnamefont{M.}~\bibnamefont{M\"ott\"onen}},
  \bibinfo{journal}{Phys. Rev. B} \textbf{\bibinfo{volume}{83}},
  \bibinfo{pages}{214518} (\bibinfo{year}{2011}).

\bibitem[{\citenamefont{Albert et~al.}(2016)\citenamefont{Albert, Shu,
  Krastanov, Shen, Liu, Yang, Schoelkopf, Mirrahimi, Devoret, and
  Jiang}}]{albert2016holonomic}
\bibinfo{author}{\bibfnamefont{V.~V.} \bibnamefont{Albert}},
  \bibinfo{author}{\bibfnamefont{C.}~\bibnamefont{Shu}},
  \bibinfo{author}{\bibfnamefont{S.}~\bibnamefont{Krastanov}},
  \bibinfo{author}{\bibfnamefont{C.}~\bibnamefont{Shen}},
  \bibinfo{author}{\bibfnamefont{R.-B.} \bibnamefont{Liu}},
  \bibinfo{author}{\bibfnamefont{Z.-B.} \bibnamefont{Yang}},
  \bibinfo{author}{\bibfnamefont{R.~J.} \bibnamefont{Schoelkopf}},
  \bibinfo{author}{\bibfnamefont{M.}~\bibnamefont{Mirrahimi}},
  \bibinfo{author}{\bibfnamefont{M.~H.} \bibnamefont{Devoret}},
  \bibnamefont{and} \bibinfo{author}{\bibfnamefont{L.}~\bibnamefont{Jiang}},
  \bibinfo{journal}{Phys. Rev. Lett.} \textbf{\bibinfo{volume}{116}},
  \bibinfo{pages}{140502} (\bibinfo{year}{2016}).

\bibitem[{\citenamefont{Aharonov and Anandan}(1987)}]{aharonov1987phase}
\bibinfo{author}{\bibfnamefont{Y.}~\bibnamefont{Aharonov}} \bibnamefont{and}
  \bibinfo{author}{\bibfnamefont{J.}~\bibnamefont{Anandan}},
  \bibinfo{journal}{Phys. Rev. Lett.} \textbf{\bibinfo{volume}{58}},
  \bibinfo{pages}{1593} (\bibinfo{year}{1987}).

\bibitem[{\citenamefont{Sj\"oqvist et~al.}(2012)\citenamefont{Sj\"oqvist, Tong,
  Andersson, Hessmo, Johansson, and Singh}}]{sjoqvist2012non}
\bibinfo{author}{\bibfnamefont{E.}~\bibnamefont{Sj\"oqvist}},
  \bibinfo{author}{\bibfnamefont{D.~M.} \bibnamefont{Tong}},
  \bibinfo{author}{\bibfnamefont{M.}~\bibnamefont{Andersson}},
  \bibinfo{author}{\bibfnamefont{B.}~\bibnamefont{Hessmo}},
  \bibinfo{author}{\bibfnamefont{M.}~\bibnamefont{Johansson}},
  \bibnamefont{and} \bibinfo{author}{\bibfnamefont{K.}~\bibnamefont{Singh}},
  \bibinfo{journal}{New J. Phys.} \textbf{\bibinfo{volume}{14}},
  \bibinfo{pages}{103035} (\bibinfo{year}{2012}).

\bibitem[{\citenamefont{Xu et~al.}(2012)\citenamefont{Xu, Zhang, Tong,
  Sj\"oqvist, and Kwek}}]{xu2012nonadiabatic}
\bibinfo{author}{\bibfnamefont{G.~F.} \bibnamefont{Xu}},
  \bibinfo{author}{\bibfnamefont{J.}~\bibnamefont{Zhang}},
  \bibinfo{author}{\bibfnamefont{D.~M.} \bibnamefont{Tong}},
  \bibinfo{author}{\bibfnamefont{E.}~\bibnamefont{Sj\"oqvist}},
  \bibnamefont{and} \bibinfo{author}{\bibfnamefont{L.~C.} \bibnamefont{Kwek}},
  \bibinfo{journal}{Phys. Rev. Lett.} \textbf{\bibinfo{volume}{109}},
  \bibinfo{pages}{170501} (\bibinfo{year}{2012}).

\bibitem[{\citenamefont{Abdumalikov et~al.}(2013)\citenamefont{Abdumalikov,
  Fink, Juliusson, Pechal, Berger, Wallraff, and
  Filipp}}]{abdumalikov2013experimental}
\bibinfo{author}{\bibfnamefont{A.~A.} \bibnamefont{Abdumalikov}},
  \bibinfo{author}{\bibfnamefont{J.~M.} \bibnamefont{Fink}},
  \bibinfo{author}{\bibfnamefont{K.}~\bibnamefont{Juliusson}},
  \bibinfo{author}{\bibfnamefont{M.}~\bibnamefont{Pechal}},
  \bibinfo{author}{\bibfnamefont{S.}~\bibnamefont{Berger}},
  \bibinfo{author}{\bibfnamefont{A.}~\bibnamefont{Wallraff}}, \bibnamefont{and}
  \bibinfo{author}{\bibfnamefont{S.}~\bibnamefont{Filipp}},
  \bibinfo{journal}{Nature~(London)} \textbf{\bibinfo{volume}{496}},
  \bibinfo{pages}{482} (\bibinfo{year}{2013}).

\bibitem[{\citenamefont{{Egger} et~al.}(2018)\citenamefont{{Egger}, {Ganzhorn},
  {Salis}, {Fuhrer}, {Mueller}, {Barkoutsos}, {Moll}, {Tavernelli}, and
  {Filipp}}}]{Egger2018Entanglement}
\bibinfo{author}{\bibfnamefont{D.~J.} \bibnamefont{{Egger}}},
  \bibinfo{author}{\bibfnamefont{M.}~\bibnamefont{{Ganzhorn}}},
  \bibinfo{author}{\bibfnamefont{G.}~\bibnamefont{{Salis}}},
  \bibinfo{author}{\bibfnamefont{A.}~\bibnamefont{{Fuhrer}}},
  \bibinfo{author}{\bibfnamefont{P.}~\bibnamefont{{Mueller}}},
  \bibinfo{author}{\bibfnamefont{P.~K.} \bibnamefont{{Barkoutsos}}},
  \bibinfo{author}{\bibfnamefont{N.}~\bibnamefont{{Moll}}},
  \bibinfo{author}{\bibfnamefont{I.}~\bibnamefont{{Tavernelli}}},
  \bibnamefont{and} \bibinfo{author}{\bibfnamefont{S.}~\bibnamefont{{Filipp}}},
  \bibinfo{journal}{arXiv: 1804.04900}.

\bibitem[{\citenamefont{Feng et~al.}(2013)\citenamefont{Feng, Xu, and
  Long}}]{Feng2013Experimental}
\bibinfo{author}{\bibfnamefont{G.}~\bibnamefont{Feng}},
  \bibinfo{author}{\bibfnamefont{G.}~\bibnamefont{Xu}}, \bibnamefont{and}
  \bibinfo{author}{\bibfnamefont{G.}~\bibnamefont{Long}},
  \bibinfo{journal}{Phys. Rev. Lett.} \textbf{\bibinfo{volume}{110}},
  \bibinfo{pages}{190501} (\bibinfo{year}{2013}).

\bibitem[{\citenamefont{Zu et~al.}(2014)\citenamefont{Zu, Wang, He, Zhang, Dai,
  Wang, and Duan}}]{zu2014experimental}
\bibinfo{author}{\bibfnamefont{C.}~\bibnamefont{Zu}},
  \bibinfo{author}{\bibfnamefont{W.}~\bibnamefont{Wang}},
  \bibinfo{author}{\bibfnamefont{L.}~\bibnamefont{He}},
  \bibinfo{author}{\bibfnamefont{W.}~\bibnamefont{Zhang}},
  \bibinfo{author}{\bibfnamefont{C.~Y.} \bibnamefont{Dai}},
  \bibinfo{author}{\bibfnamefont{F.}~\bibnamefont{Wang}}, \bibnamefont{and}
  \bibinfo{author}{\bibfnamefont{L.~M.} \bibnamefont{Duan}},
  \bibinfo{journal}{Nature~(London)} \textbf{\bibinfo{volume}{514}}, \bibinfo{pages}{72}
  (\bibinfo{year}{2014}).

\bibitem[{\citenamefont{Arroyocamejo et~al.}(2014)\citenamefont{Arroyocamejo,
  Lazariev, Hell, and Balasubramanian}}]{arroyocamejo2014room}
\bibinfo{author}{\bibfnamefont{S.}~\bibnamefont{Arroyocamejo}},
  \bibinfo{author}{\bibfnamefont{A.}~\bibnamefont{Lazariev}},
  \bibinfo{author}{\bibfnamefont{S.~W.} \bibnamefont{Hell}}, \bibnamefont{and}
  \bibinfo{author}{\bibfnamefont{G.}~\bibnamefont{Balasubramanian}},
  \bibinfo{journal}{Nat. Commun.} \textbf{\bibinfo{volume}{5}},
  \bibinfo{pages}{4870} (\bibinfo{year}{2014}).

\bibitem[{\citenamefont{Sekiguchi et~al.}(2017)\citenamefont{Sekiguchi,
  Niikura, Kuroiwa, Kano, and Kosaka}}]{yuhei2017optical}
\bibinfo{author}{\bibfnamefont{Y.}~\bibnamefont{Sekiguchi}},
  \bibinfo{author}{\bibfnamefont{N.}~\bibnamefont{Niikura}},
  \bibinfo{author}{\bibfnamefont{R.}~\bibnamefont{Kuroiwa}},
  \bibinfo{author}{\bibfnamefont{H.}~\bibnamefont{Kano}}, \bibnamefont{and}
  \bibinfo{author}{\bibfnamefont{H.}~\bibnamefont{Kosaka}},
  \bibinfo{journal}{Nature Photon.} \textbf{\bibinfo{volume}{11}},
  \bibinfo{pages}{309} (\bibinfo{year}{2017}).

\bibitem[{\citenamefont{Xu et~al.}(2015)\citenamefont{Xu, Liu, Zhao, and
  Tong}}]{xu2015nonadiabatic}
\bibinfo{author}{\bibfnamefont{G.~F.} \bibnamefont{Xu}},
  \bibinfo{author}{\bibfnamefont{C.~L.} \bibnamefont{Liu}},
  \bibinfo{author}{\bibfnamefont{P.~Z.} \bibnamefont{Zhao}}, \bibnamefont{and}
  \bibinfo{author}{\bibfnamefont{D.~M.} \bibnamefont{Tong}},
  \bibinfo{journal}{Phys. Rev. A} \textbf{\bibinfo{volume}{92}},
  \bibinfo{pages}{052302} (\bibinfo{year}{2015}).

\bibitem[{\citenamefont{Herterich and Sj\"oqvist}(2016)}]{herterich2016single}
\bibinfo{author}{\bibfnamefont{E.}~\bibnamefont{Herterich}} \bibnamefont{and}
  \bibinfo{author}{\bibfnamefont{E.}~\bibnamefont{Sj\"oqvist}},
  \bibinfo{journal}{Phys. Rev. A} \textbf{\bibinfo{volume}{94}},
  \bibinfo{pages}{052310} (\bibinfo{year}{2016}).

\bibitem[{\citenamefont{Hong et~al.}(2018)\citenamefont{Hong, Liu, Cai, Zhang,
  Hu, Wang, and Xue}}]{hong2017implementing}
\bibinfo{author}{\bibfnamefont{Z.-P.} \bibnamefont{Hong}},
  \bibinfo{author}{\bibfnamefont{B.-J.} \bibnamefont{Liu}},
  \bibinfo{author}{\bibfnamefont{J.-Q.} \bibnamefont{Cai}},
  \bibinfo{author}{\bibfnamefont{X.-D.} \bibnamefont{Zhang}},
  \bibinfo{author}{\bibfnamefont{Y.}~\bibnamefont{Hu}},
  \bibinfo{author}{\bibfnamefont{Z.~D.} \bibnamefont{Wang}}, \bibnamefont{and}
  \bibinfo{author}{\bibfnamefont{Z.-Y.} \bibnamefont{Xue}},
  \bibinfo{journal}{Phys. Rev. A} \textbf{\bibinfo{volume}{97}},
  \bibinfo{pages}{022332} (\bibinfo{year}{2018}).

\bibitem[{\citenamefont{Li et~al.}(2017)\citenamefont{Li, Liu, and
  Long}}]{li2017experimental}
\bibinfo{author}{\bibfnamefont{H.}~\bibnamefont{Li}},
  \bibinfo{author}{\bibfnamefont{Y.}~\bibnamefont{Liu}}, \bibnamefont{and}
  \bibinfo{author}{\bibfnamefont{G.}~\bibnamefont{Long}},
  \bibinfo{journal}{Sci. China-Phys. Mech. Astron.}
  \textbf{\bibinfo{volume}{60}}, \bibinfo{pages}{080311}
  (\bibinfo{year}{2017}).

\bibitem[{\citenamefont{Zhou et~al.}(2017)\citenamefont{Zhou, Jerger,
  Shkolnikov, Heremans, Burkard, and Awschalom}}]{zhou2017holonomic}
\bibinfo{author}{\bibfnamefont{B.~B.} \bibnamefont{Zhou}},
  \bibinfo{author}{\bibfnamefont{P.~C.} \bibnamefont{Jerger}},
  \bibinfo{author}{\bibfnamefont{V.~O.} \bibnamefont{Shkolnikov}},
  \bibinfo{author}{\bibfnamefont{F.~J.} \bibnamefont{Heremans}},
  \bibinfo{author}{\bibfnamefont{G.}~\bibnamefont{Burkard}}, \bibnamefont{and}
  \bibinfo{author}{\bibfnamefont{D.~D.} \bibnamefont{Awschalom}},
  \bibinfo{journal}{Phys. Rev. Lett.} \textbf{\bibinfo{volume}{119}},
  \bibinfo{pages}{140503} (\bibinfo{year}{2017}).

\bibitem[{\citenamefont{Wallraff et~al.}(2004)\citenamefont{Wallraff, Schuster,
  Blais, Frunzio, Huang, Majer, Kumar, Girvin, and Schoelkopf}}]{Wallraff}
\bibinfo{author}{\bibfnamefont{A.}~\bibnamefont{Wallraff}},
  \bibinfo{author}{\bibfnamefont{D.~I.} \bibnamefont{Schuster}},
  \bibinfo{author}{\bibfnamefont{A.}~\bibnamefont{Blais}},
  \bibinfo{author}{\bibfnamefont{L.}~\bibnamefont{Frunzio}},
  \bibinfo{author}{\bibfnamefont{R.-S.} \bibnamefont{Huang}},
  \bibinfo{author}{\bibfnamefont{J.}~\bibnamefont{Majer}},
  \bibinfo{author}{\bibfnamefont{S.}~\bibnamefont{Kumar}},
  \bibinfo{author}{\bibfnamefont{S.~M.} \bibnamefont{Girvin}},
  \bibnamefont{and} \bibinfo{author}{\bibfnamefont{R.~J.}
  \bibnamefont{Schoelkopf}}, \bibinfo{journal}{Nature~(London)}
  \textbf{\bibinfo{volume}{431}}, \bibinfo{pages}{162} (\bibinfo{year}{2004}).

\bibitem[{\citenamefont{Clarke and Wilhelm}(2008)}]{Clarke2008Superconducting}
\bibinfo{author}{\bibfnamefont{J.}~\bibnamefont{Clarke}} \bibnamefont{and}
  \bibinfo{author}{\bibfnamefont{F.~K.} \bibnamefont{Wilhelm}},
  \bibinfo{journal}{Nature~(London)} \textbf{\bibinfo{volume}{453}},
  \bibinfo{pages}{1031} (\bibinfo{year}{2008}).

\bibitem[{\citenamefont{You and Nori}(2011)}]{You2011Atomic}
\bibinfo{author}{\bibfnamefont{J.~Q.} \bibnamefont{You}} \bibnamefont{and}
  \bibinfo{author}{\bibfnamefont{F.}~\bibnamefont{Nori}},
  \bibinfo{journal}{Nature~(London)} \textbf{\bibinfo{volume}{474}},
  \bibinfo{pages}{589} (\bibinfo{year}{2011}).

\bibitem[{\citenamefont{Devoret and
  Schoelkopf}(2013)}]{devoret2013superconducting}
\bibinfo{author}{\bibfnamefont{M.~H.} \bibnamefont{Devoret}} \bibnamefont{and}
  \bibinfo{author}{\bibfnamefont{R.~J.} \bibnamefont{Schoelkopf}},
  \bibinfo{journal}{Science} \textbf{\bibinfo{volume}{339}},
  \bibinfo{pages}{1169} (\bibinfo{year}{2013}).

\bibitem[{\citenamefont{Gu et~al.}(2017)\citenamefont{Gu, Kockum, Miranowicz,
  Liu, and Nori}}]{Gu2017Microwave}
\bibinfo{author}{\bibfnamefont{X.}~\bibnamefont{Gu}},
  \bibinfo{author}{\bibfnamefont{A.~F.} \bibnamefont{Kockum}},
  \bibinfo{author}{\bibfnamefont{A.}~\bibnamefont{Miranowicz}},
  \bibinfo{author}{\bibfnamefont{Y.~X.} \bibnamefont{Liu}}, \bibnamefont{and}
  \bibinfo{author}{\bibfnamefont{F.}~\bibnamefont{Nori}},
  \bibinfo{journal}{Phys. Rep.} \textbf{\bibinfo{volume}{718-719}},
  \bibinfo{pages}{1} (\bibinfo{year}{2017}).

\bibitem[{\citenamefont{Reagor et~al.}(2013)\citenamefont{Reagor, Paik,
  Catelani, Sun, Axline, Holland, Pop, Masluk, Brecht, Frunzio
  et~al.}}]{Reagor}
\bibinfo{author}{\bibfnamefont{M.}~\bibnamefont{Reagor}},
  \bibinfo{author}{\bibfnamefont{H.}~\bibnamefont{Paik}},
  \bibinfo{author}{\bibfnamefont{G.}~\bibnamefont{Catelani}},
  \bibinfo{author}{\bibfnamefont{L.}~\bibnamefont{Sun}},
  \bibinfo{author}{\bibfnamefont{C.}~\bibnamefont{Axline}},
  \bibinfo{author}{\bibfnamefont{E.}~\bibnamefont{Holland}},
  \bibinfo{author}{\bibfnamefont{I.~M.} \bibnamefont{Pop}},
  \bibinfo{author}{\bibfnamefont{N.~A.} \bibnamefont{Masluk}},
  \bibinfo{author}{\bibfnamefont{T.}~\bibnamefont{Brecht}},
  \bibinfo{author}{\bibfnamefont{L.}~\bibnamefont{Frunzio}} \bibnamefont{\textit{et~al.}},
 \bibinfo{journal}{Appl. Phys. Lett.}
  \textbf{\bibinfo{volume}{102}}, \bibinfo{pages}{192604}
  (\bibinfo{year}{2013}).

\bibitem[{\citenamefont{Reagor et~al.}(2016)\citenamefont{Reagor, Pfaff,
  Axline, Heeres, Ofek, Sliwa, Holland, Wang, Blumoff, Chou
  et~al.}}]{Reagor2016}
\bibinfo{author}{\bibfnamefont{M.}~\bibnamefont{Reagor}},
  \bibinfo{author}{\bibfnamefont{W.}~\bibnamefont{Pfaff}},
  \bibinfo{author}{\bibfnamefont{C.}~\bibnamefont{Axline}},
  \bibinfo{author}{\bibfnamefont{R.~W.} \bibnamefont{Heeres}},
  \bibinfo{author}{\bibfnamefont{N.}~\bibnamefont{Ofek}},
  \bibinfo{author}{\bibfnamefont{K.}~\bibnamefont{Sliwa}},
  \bibinfo{author}{\bibfnamefont{E.}~\bibnamefont{Holland}},
  \bibinfo{author}{\bibfnamefont{C.}~\bibnamefont{Wang}},
  \bibinfo{author}{\bibfnamefont{J.}~\bibnamefont{Blumoff}},
  \bibinfo{author}{\bibfnamefont{K.}~\bibnamefont{Chou}} \bibnamefont{\textit{et~al.}},
  \bibinfo{journal}{Phys. Rev. B} \textbf{\bibinfo{volume}{94}},
  \bibinfo{pages}{014506} (\bibinfo{year}{2016}).

\bibitem[{\citenamefont{Leghtas et~al.}(2013)\citenamefont{Leghtas, Kirchmair,
  Vlastakis, Schoelkopf, Devoret, and Mirrahimi}}]{LeghtasPRL2013}
\bibinfo{author}{\bibfnamefont{Z.}~\bibnamefont{Leghtas}},
  \bibinfo{author}{\bibfnamefont{G.}~\bibnamefont{Kirchmair}},
  \bibinfo{author}{\bibfnamefont{B.}~\bibnamefont{Vlastakis}},
  \bibinfo{author}{\bibfnamefont{R.~J.} \bibnamefont{Schoelkopf}},
  \bibinfo{author}{\bibfnamefont{M.~H.} \bibnamefont{Devoret}},
  \bibnamefont{and}
  \bibinfo{author}{\bibfnamefont{M.}~\bibnamefont{Mirrahimi}},
  \bibinfo{journal}{Phys. Rev. Lett.} \textbf{\bibinfo{volume}{111}},
  \bibinfo{pages}{120501} (\bibinfo{year}{2013}).

\bibitem[{\citenamefont{Michael et~al.}(2016)\citenamefont{Michael, Silveri,
  Brierley, Albert, Salmilehto, Jiang, and Girvin}}]{Michael2016}
\bibinfo{author}{\bibfnamefont{M.~H.} \bibnamefont{Michael}},
  \bibinfo{author}{\bibfnamefont{M.}~\bibnamefont{Silveri}},
  \bibinfo{author}{\bibfnamefont{R.~T.} \bibnamefont{Brierley}},
  \bibinfo{author}{\bibfnamefont{V.~V.} \bibnamefont{Albert}},
  \bibinfo{author}{\bibfnamefont{J.}~\bibnamefont{Salmilehto}},
  \bibinfo{author}{\bibfnamefont{L.}~\bibnamefont{Jiang}}, \bibnamefont{and}
  \bibinfo{author}{\bibfnamefont{S.~M.} \bibnamefont{Girvin}},
  \bibinfo{journal}{Phys. Rev. X} \textbf{\bibinfo{volume}{6}},
  \bibinfo{pages}{031006} (\bibinfo{year}{2016}).

\bibitem[{Sup()}]{Supplement}
\bibinfo{journal}{{See Supplementary Material for a discussion of the experimental device and setup, system Hamiltonian, quantum process tomography, calibration of the two-photon transition and cavity-assisted transition drives, and noise-resilient feature of the holonomic gates, which includes Refs.~\cite{Reagor,Reagor2016,Raftery2017Direct,Motzoi,gambetta2011analytic,peterer2015coherence,Heeres2015,bianchetti2010control,reedthesis,thew2002qudit,
Daniel2001Measurement,Nielsen,abdumalikov2013experimental,zhu2005geometric,solinas2012on,johansson2012robustness,Berger2013Exploring,Yale2016Optical} }}.

\bibitem[{\citenamefont{Paik et~al.}(2011)\citenamefont{Paik, Schuster, Bishop,
  Kirchmair, Catelani, Sears, Johnson, Reagor, Frunzio, Glazman et~al.}}]{Paik}
\bibinfo{author}{\bibfnamefont{H.}~\bibnamefont{Paik}},
  \bibinfo{author}{\bibfnamefont{D.~I.} \bibnamefont{Schuster}},
  \bibinfo{author}{\bibfnamefont{L.~S.} \bibnamefont{Bishop}},
  \bibinfo{author}{\bibfnamefont{G.}~\bibnamefont{Kirchmair}},
  \bibinfo{author}{\bibfnamefont{G.}~\bibnamefont{Catelani}},
  \bibinfo{author}{\bibfnamefont{A.~P.} \bibnamefont{Sears}},
  \bibinfo{author}{\bibfnamefont{B.~R.} \bibnamefont{Johnson}},
  \bibinfo{author}{\bibfnamefont{M.~J.} \bibnamefont{Reagor}},
  \bibinfo{author}{\bibfnamefont{L.}~\bibnamefont{Frunzio}},
  \bibinfo{author}{\bibfnamefont{L.~I.} \bibnamefont{Glazman}} \bibnamefont{\textit{et~al.}}, 
  \bibinfo{journal}{Phys. Rev. Lett.}
  \textbf{\bibinfo{volume}{107}}, \bibinfo{pages}{240501}
  (\bibinfo{year}{2011}).

\bibitem[{\citenamefont{Vlastakis et~al.}(2013)\citenamefont{Vlastakis,
  Kirchmair, Leghtas, Nigg, Frunzio, Girvin, Mirrahimi, Devoret, and
  Schoelkopf}}]{Vlastakis}
\bibinfo{author}{\bibfnamefont{B.}~\bibnamefont{Vlastakis}},
  \bibinfo{author}{\bibfnamefont{G.}~\bibnamefont{Kirchmair}},
  \bibinfo{author}{\bibfnamefont{Z.}~\bibnamefont{Leghtas}},
  \bibinfo{author}{\bibfnamefont{S.~E.} \bibnamefont{Nigg}},
  \bibinfo{author}{\bibfnamefont{L.}~\bibnamefont{Frunzio}},
  \bibinfo{author}{\bibfnamefont{S.~M.} \bibnamefont{Girvin}},
  \bibinfo{author}{\bibfnamefont{M.}~\bibnamefont{Mirrahimi}},
  \bibinfo{author}{\bibfnamefont{M.~H.} \bibnamefont{Devoret}},
  \bibnamefont{and} \bibinfo{author}{\bibfnamefont{R.~J.}
  \bibnamefont{Schoelkopf}}, \bibinfo{journal}{Science}
  \textbf{\bibinfo{volume}{342}}, \bibinfo{pages}{607} (\bibinfo{year}{2013}).

\bibitem[{\citenamefont{Sun et~al.}(2014)\citenamefont{Sun, Petrenko, Leghtas,
  Vlastakis, Kirchmair, Sliwa, Narla, Hatridge, Shankar, Blumoff
  et~al.}}]{SunNature}
\bibinfo{author}{\bibfnamefont{L.}~\bibnamefont{Sun}},
  \bibinfo{author}{\bibfnamefont{A.}~\bibnamefont{Petrenko}},
  \bibinfo{author}{\bibfnamefont{Z.}~\bibnamefont{Leghtas}},
  \bibinfo{author}{\bibfnamefont{B.}~\bibnamefont{Vlastakis}},
  \bibinfo{author}{\bibfnamefont{G.}~\bibnamefont{Kirchmair}},
  \bibinfo{author}{\bibfnamefont{K.~M.} \bibnamefont{Sliwa}},
  \bibinfo{author}{\bibfnamefont{A.}~\bibnamefont{Narla}},
  \bibinfo{author}{\bibfnamefont{M.}~\bibnamefont{Hatridge}},
  \bibinfo{author}{\bibfnamefont{S.}~\bibnamefont{Shankar}},
  \bibinfo{author}{\bibfnamefont{J.}~\bibnamefont{Blumoff}} \bibnamefont{\textit{et~al.}},
   \bibinfo{journal}{Nature~(London)}
  \textbf{\bibinfo{volume}{511}}, \bibinfo{pages}{444} (\bibinfo{year}{2014}).

\bibitem[{\citenamefont{Liu et~al.}(2017)\citenamefont{Liu, Xu, Wang, Zheng,
  Roy, Kundu, Chand, Ranadive, Vijay, Song et~al.}}]{Liu2017}
\bibinfo{author}{\bibfnamefont{K.}~\bibnamefont{Liu}},
  \bibinfo{author}{\bibfnamefont{Y.}~\bibnamefont{Xu}},
  \bibinfo{author}{\bibfnamefont{W.}~\bibnamefont{Wang}},
  \bibinfo{author}{\bibfnamefont{S.-B.} \bibnamefont{Zheng}},
  \bibinfo{author}{\bibfnamefont{T.}~\bibnamefont{Roy}},
  \bibinfo{author}{\bibfnamefont{S.}~\bibnamefont{Kundu}},
  \bibinfo{author}{\bibfnamefont{M.}~\bibnamefont{Chand}},
  \bibinfo{author}{\bibfnamefont{A.}~\bibnamefont{Ranadive}},
  \bibinfo{author}{\bibfnamefont{R.}~\bibnamefont{Vijay}},
  \bibinfo{author}{\bibfnamefont{Y.}~\bibnamefont{Song}} \bibnamefont{\textit{et~al.}},
  \bibinfo{journal}{Sci. Adv.} \textbf{\bibinfo{volume}{3}},
  \bibinfo{pages}{e1603159} (\bibinfo{year}{2017}).

\bibitem[{\citenamefont{Hatridge et~al.}(2011)\citenamefont{Hatridge, Vijay,
  Slichter, Clarke, and Siddiqi}}]{Hatridge}
\bibinfo{author}{\bibfnamefont{M.}~\bibnamefont{Hatridge}},
  \bibinfo{author}{\bibfnamefont{R.}~\bibnamefont{Vijay}},
  \bibinfo{author}{\bibfnamefont{D.~H.} \bibnamefont{Slichter}},
  \bibinfo{author}{\bibfnamefont{J.}~\bibnamefont{Clarke}}, \bibnamefont{and}
  \bibinfo{author}{\bibfnamefont{I.}~\bibnamefont{Siddiqi}},
  \bibinfo{journal}{Phys. Rev. B} \textbf{\bibinfo{volume}{83}},
  \bibinfo{pages}{134501} (\bibinfo{year}{2011}).

\bibitem[{\citenamefont{Roy et~al.}(2015)\citenamefont{Roy, Kundu, Chand,
  Vadiraj, Ranadive, Nehra, Patankar, Aumentado, Clerk, and Vijay}}]{Roy}
\bibinfo{author}{\bibfnamefont{T.}~\bibnamefont{Roy}},
  \bibinfo{author}{\bibfnamefont{S.}~\bibnamefont{Kundu}},
  \bibinfo{author}{\bibfnamefont{M.}~\bibnamefont{Chand}},
  \bibinfo{author}{\bibfnamefont{A.~M.} \bibnamefont{Vadiraj}},
  \bibinfo{author}{\bibfnamefont{A.}~\bibnamefont{Ranadive}},
  \bibinfo{author}{\bibfnamefont{N.}~\bibnamefont{Nehra}},
  \bibinfo{author}{\bibfnamefont{M.~P.} \bibnamefont{Patankar}},
  \bibinfo{author}{\bibfnamefont{J.}~\bibnamefont{Aumentado}},
  \bibinfo{author}{\bibfnamefont{A.~A.} \bibnamefont{Clerk}}, \bibnamefont{and}
  \bibinfo{author}{\bibfnamefont{R.}~\bibnamefont{Vijay}},
  \bibinfo{journal}{Appl. Phys. Lett.} \textbf{\bibinfo{volume}{107}},
  \bibinfo{pages}{262601} (\bibinfo{year}{2015}).

\bibitem[{\citenamefont{Kamal et~al.}(2009)\citenamefont{Kamal, Marblestone,
  and Devoret}}]{Kamal}
\bibinfo{author}{\bibfnamefont{A.}~\bibnamefont{Kamal}},
  \bibinfo{author}{\bibfnamefont{A.}~\bibnamefont{Marblestone}},
  \bibnamefont{and} \bibinfo{author}{\bibfnamefont{M.~H.}
  \bibnamefont{Devoret}}, \bibinfo{journal}{Phys. Rev. B}
  \textbf{\bibinfo{volume}{79}}, \bibinfo{pages}{184301}
  (\bibinfo{year}{2009}).

\bibitem[{\citenamefont{Murch et~al.}(2013)\citenamefont{Murch, Weber, Macklin,
  and Siddiqi}}]{Murch}
\bibinfo{author}{\bibfnamefont{K.~W.} \bibnamefont{Murch}},
  \bibinfo{author}{\bibfnamefont{S.~J.} \bibnamefont{Weber}},
  \bibinfo{author}{\bibfnamefont{C.}~\bibnamefont{Macklin}}, \bibnamefont{and}
  \bibinfo{author}{\bibfnamefont{I.}~\bibnamefont{Siddiqi}},
  \bibinfo{journal}{Nature~(London)} \textbf{\bibinfo{volume}{502}},
  \bibinfo{pages}{211} (\bibinfo{year}{2013}).

\bibitem[{\citenamefont{Thew et~al.}(2002)\citenamefont{Thew, Nemoto, White,
  and Munro}}]{thew2002qudit}
\bibinfo{author}{\bibfnamefont{R.~T.} \bibnamefont{Thew}},
  \bibinfo{author}{\bibfnamefont{K.}~\bibnamefont{Nemoto}},
  \bibinfo{author}{\bibfnamefont{A.~G.} \bibnamefont{White}}, \bibnamefont{and}
  \bibinfo{author}{\bibfnamefont{W.~J.} \bibnamefont{Munro}},
  \bibinfo{journal}{Phys. Rev. A} \textbf{\bibinfo{volume}{66}},
  \bibinfo{pages}{012303} (\bibinfo{year}{2002}).

\bibitem[{\citenamefont{Bianchetti et~al.}(2010)\citenamefont{Bianchetti,
  Filipp, Baur, Fink, Lang, Steffen, Boissonneault, Blais, and
  Wallraff}}]{bianchetti2010control}
\bibinfo{author}{\bibfnamefont{R.}~\bibnamefont{Bianchetti}},
  \bibinfo{author}{\bibfnamefont{S.}~\bibnamefont{Filipp}},
  \bibinfo{author}{\bibfnamefont{M.}~\bibnamefont{Baur}},
  \bibinfo{author}{\bibfnamefont{J.~M.} \bibnamefont{Fink}},
  \bibinfo{author}{\bibfnamefont{C.}~\bibnamefont{Lang}},
  \bibinfo{author}{\bibfnamefont{L.}~\bibnamefont{Steffen}},
  \bibinfo{author}{\bibfnamefont{M.}~\bibnamefont{Boissonneault}},
  \bibinfo{author}{\bibfnamefont{A.}~\bibnamefont{Blais}}, \bibnamefont{and}
  \bibinfo{author}{\bibfnamefont{A.}~\bibnamefont{Wallraff}},
  \bibinfo{journal}{Phys. Rev. Lett.} \textbf{\bibinfo{volume}{105}},
  \bibinfo{pages}{223601} (\bibinfo{year}{2010}).

\bibitem[{\citenamefont{Weinstein et~al.}(2004)\citenamefont{Weinstein, Havel,
  Emerson, and Boulant}}]{Weinstein2004}
\bibinfo{author}{\bibfnamefont{Y.~S.} \bibnamefont{Weinstein}},
  \bibinfo{author}{\bibfnamefont{T.~F.} \bibnamefont{Havel}},
  \bibinfo{author}{\bibfnamefont{J.}~\bibnamefont{Emerson}}, \bibnamefont{and}
  \bibinfo{author}{\bibfnamefont{N.}~\bibnamefont{Boulant}},
  \bibinfo{journal}{J. Chem. Phys.} \textbf{\bibinfo{volume}{121}},
  \bibinfo{pages}{6117} (\bibinfo{year}{2004}).

\bibitem[{\citenamefont{Zhang et~al.}(2012)\citenamefont{Zhang, Laflamme, and
  Suter}}]{Zhang2012PRL}
\bibinfo{author}{\bibfnamefont{J.}~\bibnamefont{Zhang}},
  \bibinfo{author}{\bibfnamefont{R.}~\bibnamefont{Laflamme}}, \bibnamefont{and}
  \bibinfo{author}{\bibfnamefont{D.}~\bibnamefont{Suter}},
  \bibinfo{journal}{Phys. Rev. Lett.} \textbf{\bibinfo{volume}{109}},
  \bibinfo{pages}{100503} (\bibinfo{year}{2012}).

\bibitem[{\citenamefont{Knill et~al.}(2008)\citenamefont{Knill, Leibfried,
  Reichle, Britton, Blakestad, Jost, Langer, Ozeri, Seidelin, and
  Wineland}}]{knill2008randomized}
\bibinfo{author}{\bibfnamefont{E.}~\bibnamefont{Knill}},
  \bibinfo{author}{\bibfnamefont{D.}~\bibnamefont{Leibfried}},
  \bibinfo{author}{\bibfnamefont{R.}~\bibnamefont{Reichle}},
  \bibinfo{author}{\bibfnamefont{J.}~\bibnamefont{Britton}},
  \bibinfo{author}{\bibfnamefont{R.~B.} \bibnamefont{Blakestad}},
  \bibinfo{author}{\bibfnamefont{J.~D.} \bibnamefont{Jost}},
  \bibinfo{author}{\bibfnamefont{C.}~\bibnamefont{Langer}},
  \bibinfo{author}{\bibfnamefont{R.}~\bibnamefont{Ozeri}},
  \bibinfo{author}{\bibfnamefont{S.}~\bibnamefont{Seidelin}}, \bibnamefont{and}
  \bibinfo{author}{\bibfnamefont{D.~J.} \bibnamefont{Wineland}},
  \bibinfo{journal}{Phys. Rev. A} \textbf{\bibinfo{volume}{77}},
  \bibinfo{pages}{012307} (\bibinfo{year}{2008}).

\bibitem[{\citenamefont{Chow et~al.}(2009)\citenamefont{Chow, Gambetta,
  Tornberg, Koch, Bishop, Houck, Johnson, Frunzio, Girvin, and
  Schoelkopf}}]{chow2009randomized}
\bibinfo{author}{\bibfnamefont{J.~M.} \bibnamefont{Chow}},
  \bibinfo{author}{\bibfnamefont{J.~M.} \bibnamefont{Gambetta}},
  \bibinfo{author}{\bibfnamefont{L.}~\bibnamefont{Tornberg}},
  \bibinfo{author}{\bibfnamefont{J.}~\bibnamefont{Koch}},
  \bibinfo{author}{\bibfnamefont{L.~S.} \bibnamefont{Bishop}},
  \bibinfo{author}{\bibfnamefont{A.~A.} \bibnamefont{Houck}},
  \bibinfo{author}{\bibfnamefont{B.~R.} \bibnamefont{Johnson}},
  \bibinfo{author}{\bibfnamefont{L.}~\bibnamefont{Frunzio}},
  \bibinfo{author}{\bibfnamefont{S.~M.} \bibnamefont{Girvin}},
  \bibnamefont{and} \bibinfo{author}{\bibfnamefont{R.~J.}
  \bibnamefont{Schoelkopf}}, \bibinfo{journal}{Phys. Rev. Lett.}
  \textbf{\bibinfo{volume}{102}}, \bibinfo{pages}{090502}
  (\bibinfo{year}{2009}).

\bibitem[{\citenamefont{Magesan et~al.}(2012)\citenamefont{Magesan, Gambetta,
  Johnson, Ryan, Chow, Merkel, da~Silva, Keefe, Rothwell, Ohki
  et~al.}}]{magesan2012efficient}
\bibinfo{author}{\bibfnamefont{E.}~\bibnamefont{Magesan}},
  \bibinfo{author}{\bibfnamefont{J.~M.} \bibnamefont{Gambetta}},
  \bibinfo{author}{\bibfnamefont{B.~R.} \bibnamefont{Johnson}},
  \bibinfo{author}{\bibfnamefont{C.~A.} \bibnamefont{Ryan}},
  \bibinfo{author}{\bibfnamefont{J.~M.} \bibnamefont{Chow}},
  \bibinfo{author}{\bibfnamefont{S.~T.} \bibnamefont{Merkel}},
  \bibinfo{author}{\bibfnamefont{M.~P.} \bibnamefont{da~Silva}},
  \bibinfo{author}{\bibfnamefont{G.~A.} \bibnamefont{Keefe}},
  \bibinfo{author}{\bibfnamefont{M.~B.} \bibnamefont{Rothwell}},
  \bibinfo{author}{\bibfnamefont{T.~A.} \bibnamefont{Ohki}} \bibnamefont{\textit{et~al.}},
  \bibinfo{journal}{Phys. Rev. Lett.}
  \textbf{\bibinfo{volume}{109}}, \bibinfo{pages}{080505}
  (\bibinfo{year}{2012}).

\bibitem[{\citenamefont{Magesan et~al.}(2011)\citenamefont{Magesan, Gambetta,
  and Emerson}}]{magesan2011scalable}
\bibinfo{author}{\bibfnamefont{E.}~\bibnamefont{Magesan}},
  \bibinfo{author}{\bibfnamefont{J.~M.} \bibnamefont{Gambetta}},
  \bibnamefont{and} \bibinfo{author}{\bibfnamefont{J.}~\bibnamefont{Emerson}},
  \bibinfo{journal}{Phys. Rev. Lett.} \textbf{\bibinfo{volume}{106}},
  \bibinfo{pages}{180504} (\bibinfo{year}{2011}).

\bibitem[{\citenamefont{Barends et~al.}(2014)\citenamefont{Barends, Kelly,
  Megrant, Veitia, Sank, Jeffrey, White, Mutus, Fowler, Campbell
  et~al.}}]{barends2014superconducting}
\bibinfo{author}{\bibfnamefont{R.}~\bibnamefont{Barends}},
  \bibinfo{author}{\bibfnamefont{J.}~\bibnamefont{Kelly}},
  \bibinfo{author}{\bibfnamefont{A.}~\bibnamefont{Megrant}},
  \bibinfo{author}{\bibfnamefont{A.}~\bibnamefont{Veitia}},
  \bibinfo{author}{\bibfnamefont{D.}~\bibnamefont{Sank}},
  \bibinfo{author}{\bibfnamefont{E.}~\bibnamefont{Jeffrey}},
  \bibinfo{author}{\bibfnamefont{T.~C.} \bibnamefont{White}},
  \bibinfo{author}{\bibfnamefont{J.}~\bibnamefont{Mutus}},
  \bibinfo{author}{\bibfnamefont{A.~G.} \bibnamefont{Fowler}},
  \bibinfo{author}{\bibfnamefont{B.}~\bibnamefont{Campbell}} \bibnamefont{\textit{et~al.}},
  \bibinfo{journal}{Nature~(London)}
  \textbf{\bibinfo{volume}{508}}, \bibinfo{pages}{500} (\bibinfo{year}{2014}).

\bibitem[{\citenamefont{Johansson
  et~al.}(2012{\natexlab{b}})\citenamefont{Johansson, Nation, and
  Nori}}]{Johansson2012}
\bibinfo{author}{\bibfnamefont{J.~R.} \bibnamefont{Johansson}},
  \bibinfo{author}{\bibfnamefont{P.~D.} \bibnamefont{Nation}},
  \bibnamefont{and} \bibinfo{author}{\bibfnamefont{F.}~\bibnamefont{Nori}},
  \bibinfo{journal}{Comp. Phys. Comm.} \textbf{\bibinfo{volume}{183}},
  \bibinfo{pages}{1760} (\bibinfo{year}{2012}{\natexlab{b}}).

\bibitem[{\citenamefont{Johansson et~al.}(2013)\citenamefont{Johansson, Nation,
  and Nori}}]{Johansson2013}
\bibinfo{author}{\bibfnamefont{J.~R.} \bibnamefont{Johansson}},
  \bibinfo{author}{\bibfnamefont{P.~D.} \bibnamefont{Nation}},
  \bibnamefont{and} \bibinfo{author}{\bibfnamefont{F.}~\bibnamefont{Nori}},
  \bibinfo{journal}{Comp. Phys. Comm.} \textbf{\bibinfo{volume}{184}},
  \bibinfo{pages}{1234} (\bibinfo{year}{2013}).

\bibitem[{\citenamefont{Zeytino\ifmmode~\breve{g}\else \u{g}\fi{}lu
  et~al.}(2015)\citenamefont{Zeytino\ifmmode~\breve{g}\else \u{g}\fi{}lu,
  Pechal, Berger, Abdumalikov, Wallraff, and Filipp}}]{zeytinoglu2015microwave}
\bibinfo{author}{\bibfnamefont{S.}~\bibnamefont{Zeytino\ifmmode~\breve{g}\else
  \u{g}\fi{}lu}}, \bibinfo{author}{\bibfnamefont{M.}~\bibnamefont{Pechal}},
  \bibinfo{author}{\bibfnamefont{S.}~\bibnamefont{Berger}},
  \bibinfo{author}{\bibfnamefont{A.~A.} \bibnamefont{Abdumalikov}},
  \bibinfo{author}{\bibfnamefont{A.}~\bibnamefont{Wallraff}}, \bibnamefont{and}
  \bibinfo{author}{\bibfnamefont{S.}~\bibnamefont{Filipp}},
  \bibinfo{journal}{Phys. Rev. A} \textbf{\bibinfo{volume}{91}},
  \bibinfo{pages}{043846} (\bibinfo{year}{2015}).

\bibitem[{\citenamefont{Pechal et~al.}(2014)\citenamefont{Pechal, Huthmacher,
  Eichler, Zeytino\ifmmode~\breve{g}\else \u{g}\fi{}lu, Abdumalikov, Berger,
  Wallraff, and Filipp}}]{pechal2014microwave}
\bibinfo{author}{\bibfnamefont{M.}~\bibnamefont{Pechal}},
  \bibinfo{author}{\bibfnamefont{L.}~\bibnamefont{Huthmacher}},
  \bibinfo{author}{\bibfnamefont{C.}~\bibnamefont{Eichler}},
  \bibinfo{author}{\bibfnamefont{S.}~\bibnamefont{Zeytino\ifmmode~\breve{g}\else
  \u{g}\fi{}lu}}, \bibinfo{author}{\bibfnamefont{A.~A.}
  \bibnamefont{Abdumalikov}},
  \bibinfo{author}{\bibfnamefont{S.}~\bibnamefont{Berger}},
  \bibinfo{author}{\bibfnamefont{A.}~\bibnamefont{Wallraff}}, \bibnamefont{and}
  \bibinfo{author}{\bibfnamefont{S.}~\bibnamefont{Filipp}},
  \bibinfo{journal}{Phys. Rev. X} \textbf{\bibinfo{volume}{4}},
  \bibinfo{pages}{041010} (\bibinfo{year}{2014}).

\bibitem[{\citenamefont{Vogel et~al.}(1993)\citenamefont{Vogel, Akulin, and
  Schleich}}]{Vogel1993}
\bibinfo{author}{\bibfnamefont{K.}~\bibnamefont{Vogel}},
  \bibinfo{author}{\bibfnamefont{V.~M.} \bibnamefont{Akulin}},
  \bibnamefont{and} \bibinfo{author}{\bibfnamefont{W.~P.}
  \bibnamefont{Schleich}}, \bibinfo{journal}{Phys. Rev. Lett.}
  \textbf{\bibinfo{volume}{71}}, \bibinfo{pages}{1816} (\bibinfo{year}{1993}).

\bibitem[{\citenamefont{Law and Eberly}(1996)}]{Law1996}
\bibinfo{author}{\bibfnamefont{C.~K.} \bibnamefont{Law}} \bibnamefont{and}
  \bibinfo{author}{\bibfnamefont{J.~H.} \bibnamefont{Eberly}},
  \bibinfo{journal}{Phys. Rev. Lett.} \textbf{\bibinfo{volume}{76}},
  \bibinfo{pages}{1055} (\bibinfo{year}{1996}).

\bibitem[{\citenamefont{Hofheinz et~al.}(2009)\citenamefont{Hofheinz, Wang,
  Ansmann, Bialczak, Lucero, Neeley, O'Connell, Sank, Wenner, Martinis
  et~al.}}]{Hofheinz2009}
\bibinfo{author}{\bibfnamefont{M.}~\bibnamefont{Hofheinz}},
  \bibinfo{author}{\bibfnamefont{H.}~\bibnamefont{Wang}},
  \bibinfo{author}{\bibfnamefont{M.}~\bibnamefont{Ansmann}},
  \bibinfo{author}{\bibfnamefont{R.~C.} \bibnamefont{Bialczak}},
  \bibinfo{author}{\bibfnamefont{E.}~\bibnamefont{Lucero}},
  \bibinfo{author}{\bibfnamefont{M.}~\bibnamefont{Neeley}},
  \bibinfo{author}{\bibfnamefont{A.~D.} \bibnamefont{O'Connell}},
  \bibinfo{author}{\bibfnamefont{D.}~\bibnamefont{Sank}},
  \bibinfo{author}{\bibfnamefont{J.}~\bibnamefont{Wenner}},
  \bibinfo{author}{\bibfnamefont{J.~M.} \bibnamefont{Martinis}} \bibnamefont{\textit{et~al.}},
  \bibinfo{journal}{Nature~(London)}
  \textbf{\bibinfo{volume}{459}}, \bibinfo{pages}{546} (\bibinfo{year}{2009}).

\bibitem[{\citenamefont{Heeres et~al.}(2015)\citenamefont{Heeres, Vlastakis,
  Holland, Krastanov, Albert, Frunzio, Jiang, and Schoelkopf}}]{Heeres2015}
\bibinfo{author}{\bibfnamefont{R.~W.} \bibnamefont{Heeres}},
  \bibinfo{author}{\bibfnamefont{B.}~\bibnamefont{Vlastakis}},
  \bibinfo{author}{\bibfnamefont{E.}~\bibnamefont{Holland}},
  \bibinfo{author}{\bibfnamefont{S.}~\bibnamefont{Krastanov}},
  \bibinfo{author}{\bibfnamefont{V.~V.} \bibnamefont{Albert}},
  \bibinfo{author}{\bibfnamefont{L.}~\bibnamefont{Frunzio}},
  \bibinfo{author}{\bibfnamefont{L.}~\bibnamefont{Jiang}}, \bibnamefont{and}
  \bibinfo{author}{\bibfnamefont{R.~J.} \bibnamefont{Schoelkopf}},
  \bibinfo{journal}{Phys. Rev. Lett.} \textbf{\bibinfo{volume}{115}},
  \bibinfo{pages}{137002} (\bibinfo{year}{2015}).

\bibitem[{\citenamefont{Wang et~al.}(2017)\citenamefont{Wang, Hu, Xu, Liu, Ma,
  Zheng, Vijay, Song, Duan, and Sun}}]{Wang2017}
\bibinfo{author}{\bibfnamefont{W.}~\bibnamefont{Wang}},
  \bibinfo{author}{\bibfnamefont{L.}~\bibnamefont{Hu}},
  \bibinfo{author}{\bibfnamefont{Y.}~\bibnamefont{Xu}},
  \bibinfo{author}{\bibfnamefont{K.}~\bibnamefont{Liu}},
  \bibinfo{author}{\bibfnamefont{Y.}~\bibnamefont{Ma}},
  \bibinfo{author}{\bibfnamefont{S.-B.} \bibnamefont{Zheng}},
  \bibinfo{author}{\bibfnamefont{R.}~\bibnamefont{Vijay}},
  \bibinfo{author}{\bibfnamefont{Y.~P.} \bibnamefont{Song}},
  \bibinfo{author}{\bibfnamefont{L.-M.} \bibnamefont{Duan}}, \bibnamefont{and}
  \bibinfo{author}{\bibfnamefont{L.}~\bibnamefont{Sun}},
  \bibinfo{journal}{Phys. Rev. Lett.} \textbf{\bibinfo{volume}{118}},
  \bibinfo{pages}{223604} (\bibinfo{year}{2017}).

\bibitem[{\citenamefont{Caves et~al.}(1980)\citenamefont{Caves, Thorne, Drever,
  Sandberg, and Zimmermann}}]{Caves1980}
\bibinfo{author}{\bibfnamefont{C.~M.} \bibnamefont{Caves}},
  \bibinfo{author}{\bibfnamefont{K.~S.} \bibnamefont{Thorne}},
  \bibinfo{author}{\bibfnamefont{R.~W.~P.} \bibnamefont{Drever}},
  \bibinfo{author}{\bibfnamefont{V.~D.} \bibnamefont{Sandberg}},
  \bibnamefont{and}
  \bibinfo{author}{\bibfnamefont{M.}~\bibnamefont{Zimmermann}},
  \bibinfo{journal}{Rev. Mod. Phys.} \textbf{\bibinfo{volume}{52}},
  \bibinfo{pages}{341} (\bibinfo{year}{1980}).

\bibitem[{\citenamefont{{Kurpiers} et~al.}(2018)\citenamefont{{Kurpiers},
  {Magnard}, {Walter}, {Royer}, {Pechal}, {Heinsoo}, {Salath{\'e}}, {Akin},
  {Storz}, {Besse} et~al.}}]{Kurpiers2017Deterministic}
\bibinfo{author}{\bibfnamefont{P.}~\bibnamefont{{Kurpiers}}},
  \bibinfo{author}{\bibfnamefont{P.}~\bibnamefont{{Magnard}}},
  \bibinfo{author}{\bibfnamefont{T.}~\bibnamefont{{Walter}}},
  \bibinfo{author}{\bibfnamefont{B.}~\bibnamefont{{Royer}}},
  \bibinfo{author}{\bibfnamefont{M.}~\bibnamefont{{Pechal}}},
  \bibinfo{author}{\bibfnamefont{J.}~\bibnamefont{{Heinsoo}}},
  \bibinfo{author}{\bibfnamefont{Y.}~\bibnamefont{{Salath{\'e}}}},
  \bibinfo{author}{\bibfnamefont{A.}~\bibnamefont{{Akin}}},
  \bibinfo{author}{\bibfnamefont{S.}~\bibnamefont{{Storz}}},
  \bibinfo{author}{\bibfnamefont{J.-C.} \bibnamefont{{Besse}}} \bibnamefont{\textit{et~al.}},
  \bibinfo{journal}{Nature~(London)}
  \textbf{\bibinfo{volume}{558}}, \bibinfo{pages}{264} (\bibinfo{year}{2018}).

\bibitem[{\citenamefont{Rosenblum et~al.}(2018)\citenamefont{Rosenblum,
  Reinhold, Mirrahimi, Jiang, Frunzio, and Schoelkopf}}]{Rosenblum2018Fault}
\bibinfo{author}{\bibfnamefont{S.}~\bibnamefont{Rosenblum}},
  \bibinfo{author}{\bibfnamefont{P.}~\bibnamefont{Reinhold}},
  \bibinfo{author}{\bibfnamefont{M.}~\bibnamefont{Mirrahimi}},
  \bibinfo{author}{\bibfnamefont{L.}~\bibnamefont{Jiang}},
  \bibinfo{author}{\bibfnamefont{L.}~\bibnamefont{Frunzio}}, \bibnamefont{and}
  \bibinfo{author}{\bibfnamefont{R.~J.} \bibnamefont{Schoelkopf}},
  \bibinfo{journal}{Science} \textbf{\bibinfo{volume}{361}},
  \bibinfo{pages}{266} (\bibinfo{year}{2018}).

\bibitem[{\citenamefont{Ishida et~al.}(2018)\citenamefont{Ishida, Nakamura,
  Tanaka, Mishima, Kano, Kuroiwa, Sekiguchi, and Kosaka}}]{Ishida2018Universal}
\bibinfo{author}{\bibfnamefont{N.}~\bibnamefont{Ishida}},
  \bibinfo{author}{\bibfnamefont{T.}~\bibnamefont{Nakamura}},
  \bibinfo{author}{\bibfnamefont{T.}~\bibnamefont{Tanaka}},
  \bibinfo{author}{\bibfnamefont{S.}~\bibnamefont{Mishima}},
  \bibinfo{author}{\bibfnamefont{H.}~\bibnamefont{Kano}},
  \bibinfo{author}{\bibfnamefont{R.}~\bibnamefont{Kuroiwa}},
  \bibinfo{author}{\bibfnamefont{Y.}~\bibnamefont{Sekiguchi}},
  \bibnamefont{and} \bibinfo{author}{\bibfnamefont{H.}~\bibnamefont{Kosaka}},
  \bibinfo{journal}{Opt. Lett.} \textbf{\bibinfo{volume}{43}},
  \bibinfo{pages}{2380} (\bibinfo{year}{2018}).

\bibitem[{\citenamefont{{Raftery} et~al.}(2017)\citenamefont{{Raftery},
  {Vrajitoarea}, {Zhang}, {Leng}, {Srinivasan}, and
  {Houck}}}]{Raftery2017Direct}
\bibinfo{author}{\bibfnamefont{J.}~\bibnamefont{{Raftery}}},
  \bibinfo{author}{\bibfnamefont{A.}~\bibnamefont{{Vrajitoarea}}},
  \bibinfo{author}{\bibfnamefont{G.}~\bibnamefont{{Zhang}}},
  \bibinfo{author}{\bibfnamefont{Z.}~\bibnamefont{{Leng}}},
  \bibinfo{author}{\bibfnamefont{S.~J.} \bibnamefont{{Srinivasan}}},
  \bibnamefont{and} \bibinfo{author}{\bibfnamefont{A.~A.}
  \bibnamefont{{Houck}}}, \bibinfo{journal}{arXiv: 1703.00942}.

\bibitem[{\citenamefont{Motzoi et~al.}(2009)\citenamefont{Motzoi, Gambetta,
  Rebentrost, and Wilhelm}}]{Motzoi}
\bibinfo{author}{\bibfnamefont{F.}~\bibnamefont{Motzoi}},
  \bibinfo{author}{\bibfnamefont{J.~M.} \bibnamefont{Gambetta}},
  \bibinfo{author}{\bibfnamefont{P.}~\bibnamefont{Rebentrost}},
  \bibnamefont{and} \bibinfo{author}{\bibfnamefont{F.~K.}
  \bibnamefont{Wilhelm}}, \bibinfo{journal}{Phys. Rev. Lett.}
  \textbf{\bibinfo{volume}{103}}, \bibinfo{pages}{110501}
  (\bibinfo{year}{2009}).

\bibitem[{\citenamefont{Gambetta et~al.}(2011)\citenamefont{Gambetta, Motzoi,
  Merkel, and Wilhelm}}]{gambetta2011analytic}
\bibinfo{author}{\bibfnamefont{J.~M.} \bibnamefont{Gambetta}},
  \bibinfo{author}{\bibfnamefont{F.}~\bibnamefont{Motzoi}},
  \bibinfo{author}{\bibfnamefont{S.~T.} \bibnamefont{Merkel}},
  \bibnamefont{and} \bibinfo{author}{\bibfnamefont{F.~K.}
  \bibnamefont{Wilhelm}}, \bibinfo{journal}{Phys. Rev. A}
  \textbf{\bibinfo{volume}{83}}, \bibinfo{pages}{012308}
  (\bibinfo{year}{2011}).

\bibitem[{\citenamefont{Peterer et~al.}(2015)\citenamefont{Peterer, Bader, Jin,
  Yan, Kamal, Gudmundsen, Leek, Orlando, Oliver, and
  Gustavsson}}]{peterer2015coherence}
\bibinfo{author}{\bibfnamefont{M.~J.} \bibnamefont{Peterer}},
  \bibinfo{author}{\bibfnamefont{S.~J.} \bibnamefont{Bader}},
  \bibinfo{author}{\bibfnamefont{X.}~\bibnamefont{Jin}},
  \bibinfo{author}{\bibfnamefont{F.}~\bibnamefont{Yan}},
  \bibinfo{author}{\bibfnamefont{A.}~\bibnamefont{Kamal}},
  \bibinfo{author}{\bibfnamefont{T.~J.} \bibnamefont{Gudmundsen}},
  \bibinfo{author}{\bibfnamefont{P.~J.} \bibnamefont{Leek}},
  \bibinfo{author}{\bibfnamefont{T.~P.} \bibnamefont{Orlando}},
  \bibinfo{author}{\bibfnamefont{W.~D.} \bibnamefont{Oliver}},
  \bibnamefont{and}
  \bibinfo{author}{\bibfnamefont{S.}~\bibnamefont{Gustavsson}},
  \bibinfo{journal}{Phys. Rev. Lett.} \textbf{\bibinfo{volume}{114}},
  \bibinfo{pages}{010501} (\bibinfo{year}{2015}).

\bibitem[{\citenamefont{Reed}(2013)}]{reedthesis}
\bibinfo{author}{\bibfnamefont{M.~D.} \bibnamefont{Reed}}, Ph.D. thesis,
  \bibinfo{school}{Yale University} (\bibinfo{year}{2013}).

\bibitem[{\citenamefont{James et~al.}(2001)\citenamefont{James, Kwiat, Munro,
  and White}}]{Daniel2001Measurement}
\bibinfo{author}{\bibfnamefont{D.~F.~V.} \bibnamefont{James}},
  \bibinfo{author}{\bibfnamefont{P.~G.} \bibnamefont{Kwiat}},
  \bibinfo{author}{\bibfnamefont{W.~J.} \bibnamefont{Munro}}, \bibnamefont{and}
  \bibinfo{author}{\bibfnamefont{A.~G.} \bibnamefont{White}},
  \bibinfo{journal}{Phys. Rev. A} \textbf{\bibinfo{volume}{64}},
  \bibinfo{pages}{052312} (\bibinfo{year}{2001}).

\bibitem[{\citenamefont{Nielsen and Chuang}(2000)}]{Nielsen}
\bibinfo{author}{\bibfnamefont{M.~A.} \bibnamefont{Nielsen}} \bibnamefont{and}
  \bibinfo{author}{\bibfnamefont{I.~L.} \bibnamefont{Chuang}},
  \emph{\bibinfo{title}{Quantum Computation and Quantum Information}}
  (\bibinfo{publisher}{Cambridge Univ. Press}, \bibinfo{year}{2000}).


\end{thebibliography}

\end{document}

\title{Supplementary Material for ``Single-loop realization of arbitrary non-adiabatic holonomic single-qubit quantum gates in a superconducting circuit"}

\author{Y. Xu}
\author{W. Cai}
\author{Y. Ma}
\author{X. Mu}
\author{L. Hu}
\affiliation{Center for Quantum Information, Institute for Interdisciplinary Information
Sciences, Tsinghua University, Beijing 100084, China}

\author{Tao Chen}
\affiliation{Guangdong Provincial Key Laboratory of Quantum Engineering and Quantum Materials, and School of Physics\\ and Telecommunication Engineering, South China Normal University, Guangzhou 510006, China}

\author{H. Wang}
\author{Y.P. Song}
\affiliation{Center for Quantum Information, Institute for Interdisciplinary Information
Sciences, Tsinghua University, Beijing 100084, China}

\author{Zheng-Yuan Xue} \email{zyxue83@163.com}
\affiliation{Guangdong Provincial Key Laboratory of Quantum Engineering and Quantum Materials, and School of Physics\\ and Telecommunication Engineering, South China Normal University, Guangzhou 510006, China}

\author{Zhang-qi Yin}\email{yinzhangqi@tsinghua.edu.cn}

\author{L.~Sun}
\email{luyansun@tsinghua.edu.cn}
\affiliation{Center for Quantum Information, Institute for Interdisciplinary Information
Sciences, Tsinghua University, Beijing 100084, China}

%\date{\today}
%\email{luyansun@tsinghua.edu.cn}
\pacs{} \maketitle

\section{Experimental Device}
\begin{figure}[htb]
\includegraphics{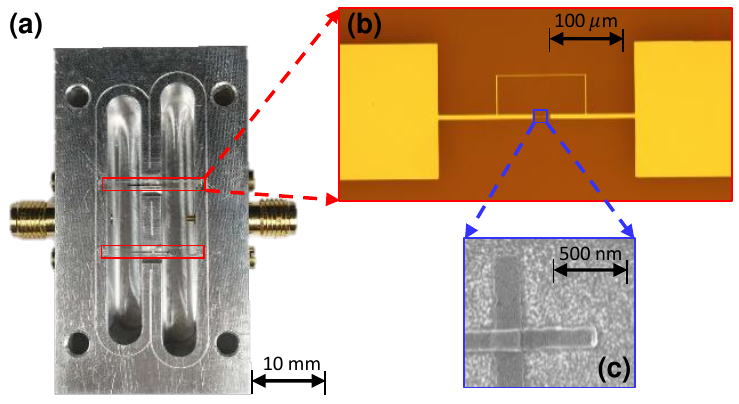}
\caption{The experimental device. (a) Optical image of the two 3D Al microwave cavities, housing two chips (depicted in red boxes). Each chip contains a superconducting transmon qubit. (b) Optical image of the qubit with a Josephson junction (depicted in a blue box) connected to two antenna pads. The thin line across the junction is used to protect the qubit during fabrication and is cut open before installing into the cavities. (c) Scanning electron microscope image of the $\mathrm{Al/AlO_x/Al}$ Josephson junction fabricated with a double-angle evaporation.}
\label{fig:device}
\end{figure}

Our experimental device contains two superconducting transmon qubits in two trenches strongly coupled to two three-dimensional (3D) waveguide cavities, as shown in Fig.~\ref{fig:device}. The qubits, each of which consists of a Josephson junction connected to two antenna pads with different lengths, are fabricated on two separate $c$-plane sapphire chips with a double-angle evaporation of aluminum after a single electron-beam lithography step. The two qubits are separated by 10~mm to suppress their direct crosstalk. The two cavities are machined from a block of high purity (5N5) aluminum and are chemically etched for a better coherence quality~\cite{Reagor,Reagor2016}. One cavity is for a fast and high-fidelity joint readout of the two qubits, while the other one is for storage and manipulation of the photonic states. The input and output couplings of the cavities can be adjusted by tuning the lengths of the input and output coupler pins.

%Our experimental device consists of two superconducting transmon qubits in two separate trenches strongly coupled to two three-dimensional (3D) waveguide cavities, as shown in Fig.~\ref{fig:device}. The two cavities are machined from a block of high purity (5N5) aluminum and are chemically etched for a better coherence quality~\cite{Reagor,Reagor2016}. One cavity is for a fast and high-fidelity joint readout of the two-qubit states, while the other one is for storage and manipulation of the photonic states. Each of the two trenches holds a 1.2~mm$\times$15.7~mm $c$-plane sapphire chip with a thickness of 430~$\mu$m, on which a transmon qubit, consisting of a Josephson junction connected to two antenna pads with different lengths, is fabricated with a double-angle evaporation of aluminum after a single electron-beam lithography step. The two chips are separated by 10~mm to suppress the crosstalk between the two qubits. The input and output couplings of the cavities can be adjusted by tuning the lengths of the input and output coupler pins.

\section{Experimental setup}
\begin{figure*}[tbp]
\includegraphics{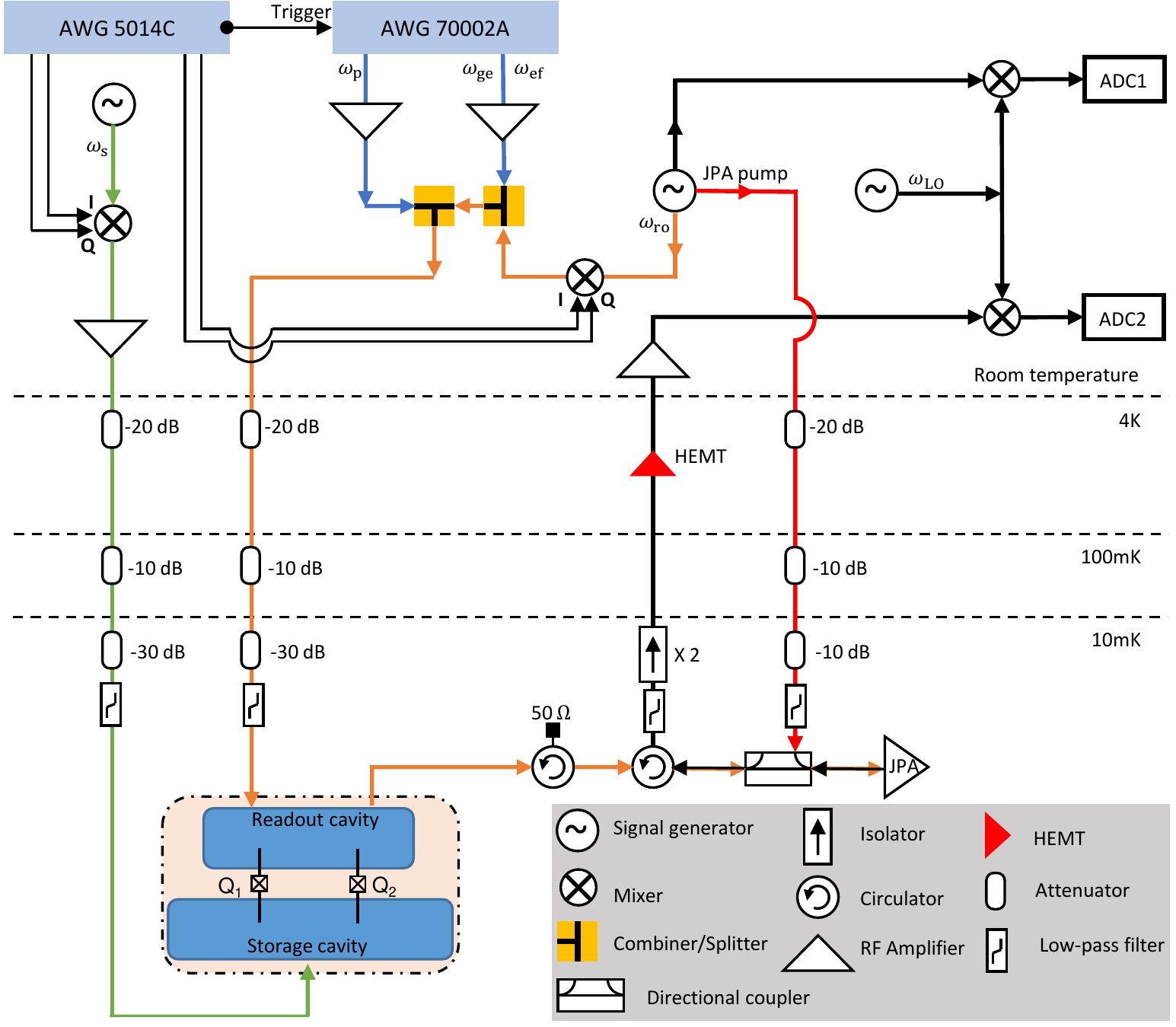}
\caption{Schematic of the full wiring of the experimental setup.  }
\label{fig:setup}
\end{figure*}

The experimental device is installed inside a magnetic shield and cooled down to $T\approx 10$~mK in a cryogen-free dilution refrigerator. The full wiring diagram is shown in Fig.~\ref{fig:setup}. The qubit control pulses are generated directly from Tektronix arbitrary waveform generator (AWG) 70002A benefiting from its large bandwidth and sampling rate~\cite{Raftery2017Direct}. The pulses for the initial state preparations and the pre-rotations before measurement have a truncated Gaussian envelope with a width of $4\sigma=40~$ns. The technique of ``derivative removal by adiabatic gate" (DRAG) is applied to both $\ket e \leftrightarrow \ket f$ and $\ket g \leftrightarrow \ket e$ transition drives in order to remove the leakage to the undesired energy levels~\cite{Motzoi,gambetta2011analytic}. The storage cavity drives are generated by IQ modulations with two analog channels of a Tektronix AWG 5014C for arbitrary cavity controls. Another two analog channels of the AWG 5014C are used to modulate the readout pulse. The readout signal is amplified by a Josephson parameter amplifier (JPA) at base temperature, a high electron mobility transistor (HEMT) at 4K stage, and a standard commercial amplifier at room temperature. Finally, the readout signal is mixed down to 50~MHz with a local oscillator (LO) before being digitized and recorded by the analog-to-digital converters (ADC).

\section{System Hamiltonian}
In our device, two transmon qubits are dispersively coupled to two 3D cavity modes. Each transmon has a large anharmonicity and is considered as a three-level artificial atom. The whole system can be described by the following Hamiltonian
\begin{eqnarray} \label{h}
\mathcal{H}/\hbar &=& \omega_\mathrm{r}\left(a_\mathrm{r}^{\dagger}a_\mathrm{r}+1/2\right) + \omega_\mathrm{s}\left(a_\mathrm{s}^{\dagger}a_\mathrm{s}+1/2\right)\notag\\
&+& \omega_{\mathrm{ge1}}\ket{e_1}\bra{e_1} + \left(\omega_{\mathrm{ge1}}+\omega_{\mathrm{ef1}}\right)\ket{f_1}\bra{f_1}\notag\\
&+& \omega_{\mathrm{ge2}}\ket{e_2}\bra{e_2} + \left(\omega_{\mathrm{ge2}}+\omega_{\mathrm{ef2}}\right)\ket{f_2}\bra{f_2}\notag\\
&-& \chi_{\mathrm{rq1}}^{\mathrm{ge}}\ket{e_1}\bra{e_1}a_\mathrm{r}^{\dagger}a_\mathrm{r} - \left(\chi_{\mathrm{rq1}}^{\mathrm{ge}}+\chi_{\mathrm{rq1}}^{\mathrm{ef}}\right)\ket{f_1}\bra{f_1}a_\mathrm{r}^{\dagger}a_\mathrm{r}\notag\\
&-& \chi_{\mathrm{rq2}}^{\mathrm{ge}}\ket{e_2}\bra{e_2}a_\mathrm{r}^{\dagger}a_\mathrm{r} - \left(\chi_{\mathrm{rq2}}^{\mathrm{ge}}+\chi_{\mathrm{rq2}}^{\mathrm{ef}}\right)\ket{f_2}\bra{f_2}a_\mathrm{r}^{\dagger}a_\mathrm{r}\notag\\
&-& \chi_{\mathrm{sq1}}^{\mathrm{ge}}\ket{e_1}\bra{e_1}a_\mathrm{s}^{\dagger}a_\mathrm{s} - \left(\chi_{\mathrm{sq1}}^{\mathrm{ge}}+\chi_{\mathrm{sq1}}^{\mathrm{ef}}\right)\ket{f_1}\bra{f_1}a_\mathrm{s}^{\dagger}a_\mathrm{s}\notag\\
&-& \chi_{\mathrm{sq2}}^{\mathrm{ge}}\ket{e_2}\bra{e_2}a_\mathrm{s}^{\dagger}a_\mathrm{s} - \left(\chi_{\mathrm{sq2}}^{\mathrm{ge}}+\chi_{\mathrm{sq2}}^{\mathrm{ef}}\right)\ket{f_2}\bra{f_2}a_\mathrm{s}^{\dagger}a_\mathrm{s}\notag\\
&-& \frac{K_\mathrm{r}}{2} a_\mathrm{r}^{\dagger}a_\mathrm{r}^{\dagger}a_\mathrm{r}a_\mathrm{r} - \frac{K_\mathrm{s}}{2} a_\mathrm{s}^{\dagger}a_\mathrm{s}^{\dagger}a_\mathrm{s}a_\mathrm{s},
% - \chi_{\mathrm{rs}} a_\mathrm{r}^{\dagger}a_\mathrm{r}a_\mathrm{s}^{\dagger}a_\mathrm{s}
\end{eqnarray}
where $\omega_\mathrm{r,s}$ are the readout and the storage cavity frequency, respectively; $a_\mathrm{r,s}$ are the corresponding ladder operators; $\omega_{\mathrm{ge}i}$ and $\omega_{\mathrm{ef}i}$ are the transition frequencies among the lowest three levels \{$\ket{g_i}$, $\ket{e_i}$, $\ket{f_i}$\} of the $i$-$\mathrm{th}$ qubit; $\chi$'s are the corresponding dispersive interactions between the two qubits and the two cavities; and $K_\mathrm{r,s}$ are the self-Kerr of the readout and the storage cavity, respectively. All the relevant parameters in the Hamiltonian are listed in Table~\ref{TableS1}. In our joint readout of the two qubits, since we only need to distinguish the $\ket{gg}$ state (both qubits in the ground state) from all others, $\chi_{\mathrm{rq1}}^{\mathrm{ef}}$, $\chi_{\mathrm{rq2}}^{\mathrm{ef}}$, and $K_\mathrm{r}$ are irrelevant and thus not measured directly.

\begin{table}[htb]
\caption{Hamiltonian parameters. $\chi_{\mathrm{rq1}}^{\mathrm{ef}}$, $\chi_{\mathrm{rq2}}^{\mathrm{ef}}$, and $K_\mathrm{r}$ are irrelevant for the joint readout and thus not measured directly. The numbers 1 and 2 in the subscript correspond to qubits $Q_1$ and $Q_2$ respectively.}
\begin{tabular}{p{2cm}<{\centering} p{2cm}<{\centering} |p{2cm}<{\centering} p{2cm}<{\centering} }
  \hline
  Terms ($/2\pi$) & Measured & Terms ($/2\pi$) & Measured \\\hline
  &&&\\[-0.9em]
  $\omega_\mathrm{r}$ & 8.540 GHz & $\omega_\mathrm{s}$ & 7.614 GHz \\
  &&&\\[-0.9em]
  $\omega_{\mathrm{ge1}}$ & 5.036 GHz & $\omega_{\mathrm{ge2}}$ & 5.605 GHz \\
  &&&\\[-0.9em]
  $\omega_{\mathrm{ef1}}$ & 4.782 GHz & $\omega_{\mathrm{ef2}}$ & 5.367 GHz \\
  &&&\\[-0.9em]
  $\chi_{\mathrm{rq1}}^{\mathrm{ge}}$ & 2.230 MHz & $\chi_{\mathrm{sq1}}^{\mathrm{ge}}$ & 0.942 MHz \\
  &&&\\[-0.9em]
  $\chi_{\mathrm{rq1}}^{\mathrm{ef}}$ & - & $\chi_{\mathrm{sq1}}^{\mathrm{ef}}$ & 0.843 MHz \\
  &&&\\[-0.9em]
  $\chi_{\mathrm{rq2}}^{\mathrm{ge}}$ & 3.00 MHz & $\chi_{\mathrm{sq2}}^{\mathrm{ge}}$ & 1.436 MHz \\
  &&&\\[-0.9em]
  $\chi_{\mathrm{rq2}}^{\mathrm{ef}}$ & - & $\chi_{\mathrm{sq2}}^{\mathrm{ef}}$ & 1.193 MHz \\
  &&&\\[-0.9em]
  $K_\mathrm{r}$ & - & $K_\mathrm{s}$ & 3.7 kHz \\
%  &&&\\[-0.9em]
%  $\chi_{\mathrm{rs}}$ & - &  &  \\
  \hline
\end{tabular}\vspace{-6pt}
\label{TableS1}
\end{table}

\begin{figure}[b]
\includegraphics{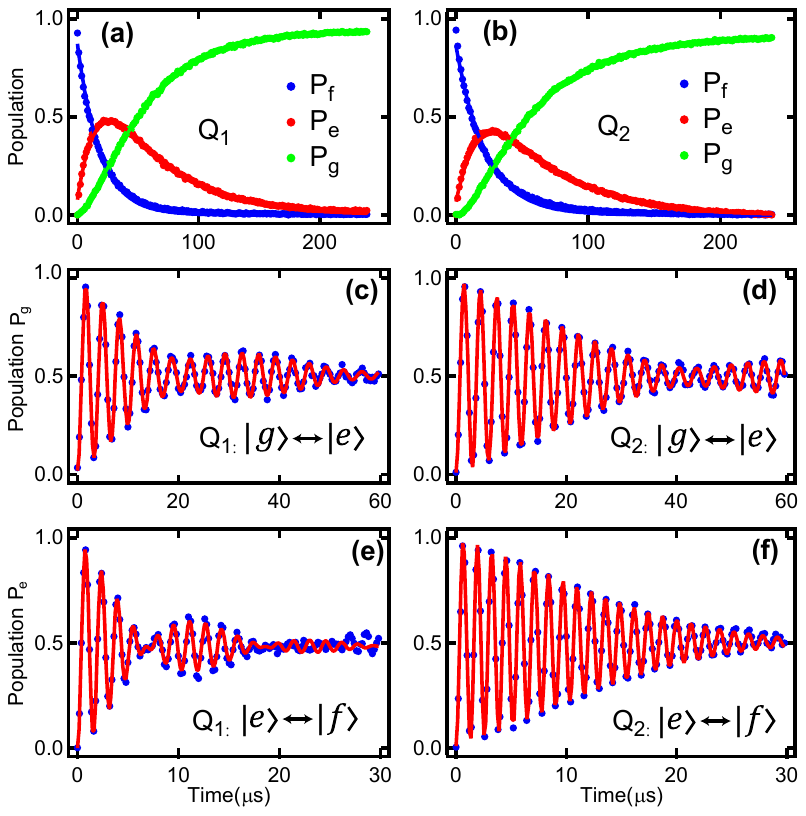}
\caption{Coherence properties of the three-level transmon qubit. (a) and (b) Population decay curves with the transmon qubit initial in $\ket f$ state for qubits $Q_1$ and $Q_2$, respectively. The solid lines are global fittings based on the rate equation. (c) and (d) Ramsey oscillation experiments between $\ket g$ and $\ket e$ for qubits $Q_1$ and $Q_2$, respectively. (e) and (f) Ramsey oscillation experiments between $\ket e$ and $\ket f$ for qubits $Q_1$ and $Q_2$, respectively. Here, a smooth background has been subtracted from the Ramsey fringes to discard the energy decay to $\ket g$. All the Ramsey oscillation fringes are fitted with the exponentially damped double sinusoidal function (red solid lines) to obtain $T_2^*$ between $\ket g$ and $\ket e$ and between $\ket e$ and $\ket f$.}
\label{fig:coherence}
\end{figure}

The coherence properties of the two qubits and the two cavities are also experimentally measured and listed in Table~\ref{TableS2}. The relaxation times of each transmon are obtained by measuring the free evolutions of the populations of the three levels $P_\mathrm{g}$, $P_\mathrm{e}$, and $P_\mathrm{f}$ with an initial $\ket{f}$ state, following the technique described in Ref.~\cite{peterer2015coherence}. $P_\mathrm{e}$ and $P_\mathrm{f}$ are measured by mapping them onto the population of $\ket{g}$ through a $\pi$ pulse and two sequential $\pi$ pulses, respectively. The experimental results are shown in Figs.~\ref{fig:coherence}a and \ref{fig:coherence}b. The decay curves are globally fitted with the rate equation $d\vec{p}/dt = \Gamma \cdot \vec{p}$, where $\vec{p}=\left(P_\mathrm{g}, P_\mathrm{e}, P_\mathrm{f}\right)^T$, and the decay rate matrix $\Gamma$ is
\begin{eqnarray}\label{gamma}
\Gamma&=&\left(\begin{array}{ccc}
 0&\Gamma_{\mathrm{eg}} &\Gamma_{\mathrm{fg}} \\
 0&-\Gamma_{\mathrm{eg}}  &\Gamma_{\mathrm{fe}} \\
 0&0  &-\left(\Gamma_{\mathrm{fe}}+\Gamma_{\mathrm{fg}}\right) \\
                        \end{array}\right),
\end{eqnarray}
where we ignore the negligible upward transition rates and only include the downward transition rates $\Gamma_{\mathrm{eg}}$, $\Gamma_{\mathrm{fe}}$, and $\Gamma_{\mathrm{fg}}$. Since the non-sequential decay rate $\Gamma_{\mathrm{fg}}$ is much slower than the sequential decay rates $\Gamma_{\mathrm{eg}}$ and $\Gamma_{\mathrm{fe}}$, the corresponding coherence times $1/\Gamma_{\mathrm{eg}}$ and $1/\Gamma_{\mathrm{fe}}$ of both qubits are listed as $T_1$ in Table~\ref{TableS2}.

The dephasing rates between $\ket g$ and $\ket e$ and between $\ket e$ and $\ket f$ of each qubit are measured with Ramsey interference experiments and the results are shown in Figs.~\ref{fig:coherence}c-f. The Ramsey fringes are fitted with an exponentially damped double sinusoidal function~\cite{peterer2015coherence} $y=y_0+e^{-t/T_2^*}\left[ A_1\cos(2\pi f_1t+\varphi_1)+A_2\cos(2\pi f_2t+\varphi_2) \right]$ and the extracted $T_2^*$ are also listed in Table~\ref{TableS2}. The coherence times $T_1$ and $T_2^*$ of the storage cavity are measured through the relaxation of Fock state $\ket 1$ and the dephasing of $\left(\ket 0 + \ket 1\right)/\sqrt{2}$, respectively~\cite{Reagor2016}. Both initial states are generated with selective number-dependent arbitrary phase gates~\cite{Heeres2015}.

\begin{table}[tb]
\caption{Coherence properties of the system.}
\begin{tabular}{p{2.6cm}<{\centering} p{2.6cm}<{\centering} p{2.6cm}<{\centering} }
  \hline
  Modes & $T_1$ & $T_2^*$ \\\hline
  readout cavity & 84~ns & -\\
  storage cavity & 135~$\mu$s & 193~$\mu$s\\
  $Q_1$ $\ket g \leftrightarrow \ket e$ & 45.6~$\mu$s & 24.4~$\mu$s  \\
  $Q_1$ $\ket e \leftrightarrow \ket f$ & 20.3~$\mu$s & 8.3~$\mu$s  \\
  $Q_2$ $\ket g \leftrightarrow \ket e$ & 42.2~$\mu$s & 44.0~$\mu$s  \\
  $Q_2$ $\ket e \leftrightarrow \ket f$ & 24.9~$\mu$s & 13.6~$\mu$s  \\
  \hline
\end{tabular}
\label{TableS2}
\end{table}

\section{Quantum process tomography}
\begin{figure}[tbp]
\includegraphics[width=9cm]{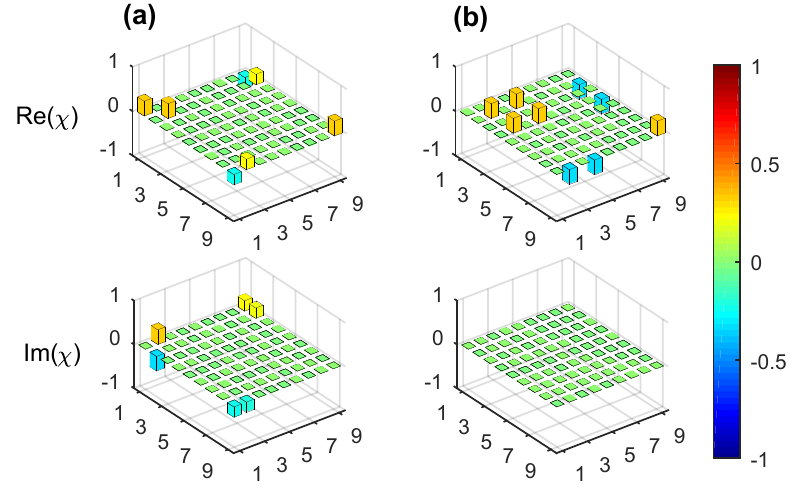}
\caption{Experimental full process matrices $\chi_{\mathrm{exp}}$. (a) and (b) Bar charts of the real and imaginary parts of the holonomic gates $X_{\pi/2}=U_1(\pi/2,\pi/2,0)$ and $H=U_1(\pi/4,\pi,0)$ respectively. The numbers in the $x$ and $y$ axes correspond to the operators in the basis set \{$I_{\mathrm{gf}}$, $\sigma_{\mathrm{gf}}^x$, $-i\sigma_{\mathrm{gf}}^y$, $\sigma_{\mathrm{gf}}^z$, $\sigma_{\mathrm{ge}}^x$, $-i\sigma_{\mathrm{ge}}^y$, $\sigma_{\mathrm{ef}}^x$, $-i\sigma_{\mathrm{ef}}^y$, $I_\mathrm{e}$\}. Solid black outlines are the corresponding ideal process matrices $\chi_{\mathrm{th}}$.}
\label{fig:FullProcessChi}
\end{figure}

The holonomic quantum gates are characterized by a full quantum process tomography of the three-level system. We first initialize the three-level transmon qubit with the following nine states \{$\ket g$, $\ket e$, $\ket f$, $\left(\ket g+\ket e \right)/\sqrt{2}$, $\left(\ket e+\ket f \right)/\sqrt{2}$, $\left(\ket g+\ket f \right)/\sqrt{2}$, $\left(\ket g-i\ket e \right)/\sqrt{2}$, $\left(\ket e-i\ket f \right)/\sqrt{2}$, $\left(\ket g-i\ket f \right)/\sqrt{2}$\}, then apply the holonomic gates, and finally perform state tomography measurements of the final states. The state tomography measurement requires nine pre-rotations to reconstruct the density matrix of the three-level qubit: \{$I$, $X_{\pi/2}^{\mathrm{ge}}$, $Y_{\pi/2}^{\mathrm{ge}}$, $X_{\pi}^{\mathrm{ge}}$, $X_{\pi/2}^{\mathrm{ge}}X_{\pi}^{\mathrm{ef}}$, $Y_{\pi/2}^{\mathrm{ge}}X_{\pi}^{\mathrm{ef}}$, $X_{\pi}^{\mathrm{ge}}X_{\pi/2}^{\mathrm{ef}}$, $X_{\pi}^{\mathrm{ge}}Y_{\pi/2}^{\mathrm{ef}}$, $X_{\pi}^{\mathrm{ge}}X_{\pi}^{\mathrm{ef}}$\}~\cite{bianchetti2010control}, where the rotation operators are read from the right to left. The measurements give the result $\langle M_k \rangle = \texttt{Tr}\left(\rho U_k^{\dagger} M_I U_k \right)$ for each pre-rotation $U_{k}$ with $k=0,1,2...8$, where $M_I=\ket{g}\bra{g}=\beta_a\lambda_0 + \beta_b\lambda_3 + \beta_c\lambda_8$ is the measurement operator~\cite{reedthesis}, $\lambda_i$ with $i=0,1,2,...8$ are the GellMann operators for a three-level qubit~\cite{thew2002qudit}, and ($\beta_a$, $\beta_b$, $\beta_c$) are the measurement coefficients which can be calibrated by preparing the qubit in $\ket g$, $\ket e$, and $\ket f$, respectively~\cite{reedthesis}. The density matrix of the three-level qubit state can then be reconstructed by the maximum likelihood estimation method~\cite{Daniel2001Measurement}. With the nine initial states $\rho_i$, the experimental process matrix $\chi_{\mathrm{exp}}$ can be extracted from the nine corresponding final states $\rho_f$ through $\rho_f=\sum_{m,n}{\chi_{mn} E_m\rho_i E_n^{\dagger}}$~\cite{Nielsen}, where the full set of nine orthogonal basis operators is chosen as \{$I_{\mathrm{gf}}$, $\sigma_{\mathrm{gf}}^x$, $-i\sigma_{\mathrm{gf}}^y$, $\sigma_{\mathrm{gf}}^z$, $\sigma_{\mathrm{ge}}^x$, $-i\sigma_{\mathrm{ge}}^y$, $\sigma_{\mathrm{ef}}^x$, $-i\sigma_{\mathrm{ef}}^y$, $I_\mathrm{e}$\}~\cite{abdumalikov2013experimental}, where $\sigma_{mn}$ are the Pauli operators acting on the $m$ and $n$ energy levels, $I_{\mathrm{gf}}=\ket{g}\bra{g}+\ket{f}\bra{f}$, and $I_\mathrm{e}=\ket{e}\bra{e}$. The full process matrices $\chi_{\mathrm{exp}}$ of the holonomic gates $X_{\pi/2}=U_1(\pi/2,\pi/2,0)$ and $H=U_1(\pi/4,\pi,0)$ on the three-level transmon qubit are shown in Fig.~\ref{fig:FullProcessChi}.

For the holonomic gates on the transmon qubit, the state $\ket e$ serves as an auxiliary state. Therefore, we have calculated the reduced process matrix $\chi_\mathrm{r}$ which describes the process only involving $\ket g$ and $\ket f$ and ignores any operators acting on the auxiliary state. In order to compare with the process acting on a two-level system, the reduced process matrix $\chi_\mathrm{r}$ is obtained by a normalization factor of 3/2. And the basis operators of the reduced process matrix are \{$I_{\mathrm{gf}}$, $\sigma_{\mathrm{gf}}^x$, $-i\sigma_{\mathrm{gf}}^y$, $\sigma_{\mathrm{gf}}^z$\}, or simply \{$I$, $X$, $-iY$, $Z$\} as in the main text.

For the holonomic gates on the cavity Fock states, only four initial states \{$\ket g$, $\ket f$, $\left(\ket g + \ket f \right)/\sqrt{2}$, $\left(\ket g -i \ket f \right)/\sqrt{2}$\} are prepared with qubit $Q_2$, and then they are mapped onto Fock states \{$\ket 0$, $\ket 1$, $\left(\ket 0 + \ket 1 \right)/\sqrt{2}$, $\left(\ket 0 -i \ket 1 \right)/\sqrt{2}$\} in the cavity with the encoding process. After performing the holonomic gates on these Fock states, the cavity states are mapped back to qubit $Q_2$ with the decoding process and finally the state tomography of qubit $Q_2$ is performed.

\section{Calibration of the two-photon transition and cavity-assisted Raman transition drives}
To implement the holonomic gates on the cavity Fock states, the effective coupling strengths and the Stark-shifts of the two-photon transition and the cavity-assisted Raman transition drives are critical and thus fully calibrated.

\begin{figure}[htb]
\includegraphics{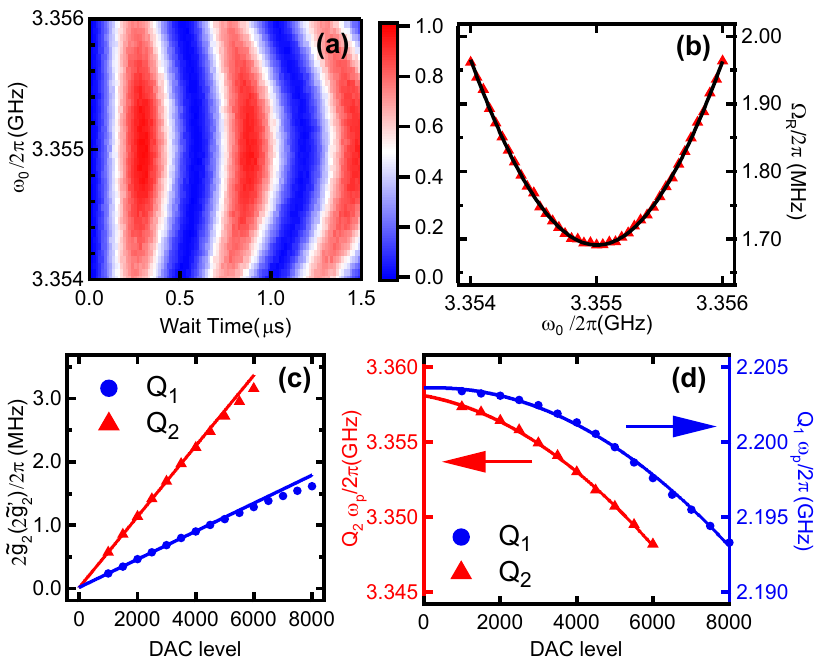}
\caption{Calibration of the cavity-assisted Raman transition drives. (a) Rabi oscillations of $\ket{0f}$ and $\ket{1g}$ at $\mathrm{DAC~level}=3000$ but with different drive durations and different drive frequencies produce a chevron pattern. Here the joint state notation denotes $\ket{\mathrm{cavity}, Q_2}$. (b) Rabi frequency $\Omega_\mathrm{R}$ of the oscillation between $\ket{0f}$ and $\ket{1g}$ obtained from (a) as a function of the drive frequency. Solid red line is a fit of the data, giving both the resonant frequency $\omega_\mathrm{p}$ and the effective coupling strength $\tilde{g}'_2$ for $Q_2$. (c) Effective coupling strengths $\tilde{g}_2$ and $\tilde{g}'_2$ for both qubits $Q_1$ and $Q_2$ as a function of the drive pulse amplitude (DAC level). The relationship between $\tilde{g}_2$ ($\tilde{g}'_2$) and the drive amplitude slightly deviates from a line at large drive strengths due to the saturation of the room temperature amplifier. (d) The resonant frequency $\omega_\mathrm{p}$ of the cavity-assisted Raman transition as a function of the drive pulse amplitude (DAC level) for both qubits $Q_1$ and $Q_2$. Solid lines are a quadratic fit of the corresponding data.}
\label{fig:RedSideBand}
\end{figure}

The cavity-assisted Raman transition drives for both qubits $Q_1$ and $Q_2$ are calibrated separately as follows, since they are used for the implementation of the holonomic gates and the encoding/decoding processes, respectively. One of the qubits is initialized in $\ket f$ state by two sequential $\pi$ pulses while the other one remains in $\ket{g}$ state during the calibration. A microwave pulse at frequency $\omega_\mathrm{0}$ with a square envelope at a fixed digital-to-analog converter (DAC) level is then applied. The coherent oscillations between the corresponding $\ket{0f}$ and $\ket{1g}$ as a function of the square pulse width and the drive frequency $\omega_\mathrm{0}$ produce a chevron pattern as shown in Fig.~\ref{fig:RedSideBand}a for $Q_2$. From a sinusoidal fit of the coherent oscillations, we obtain the Rabi frequency $\Omega_\mathrm{R}$ as a function of the drive frequency as shown in Fig.~\ref{fig:RedSideBand}b. The solid red line is fitted with the function $\Omega_\mathrm{R} = \sqrt{\Delta^2 + \left(2\tilde{g}'_2\right)^2}$, where $\Delta=\omega_\mathrm{p}-\omega_0$, giving the resonant frequency $\omega_\mathrm{p}$ and the effective coupling strength $\tilde{g}'_2$ for $Q_2$. Note that we choose $\tilde{g}_2$ ($\tilde{g}'_2$) for $Q_1$ ($Q_2$) to be consistent with the notation in the main text. Repeats of the same experiment at different amplitudes (DAC level) of the drive pulse give a calibration of the resonant frequency $\omega_\mathrm{p}$ and the effective coupling strength $\tilde{g}_2$ ($\tilde{g}_2'$) as a function of the DAC level. The results for both qubits are shown in Figs.~\ref{fig:RedSideBand}c and \ref{fig:RedSideBand}d, respectively.

The cavity-assisted Raman transition for qubit $Q_2$ is used for the encoding and decoding processes, and we choose $\mathrm{DAC~level}=3000$ for the drive pulse which corresponds to an effective coupling strength $\tilde{g}'_2/2\pi=0.845$~MHz. Therefore, a 294~ns square pulse with gradual ramp-up and ramp-down edges (10~ns for each) can provide a fast encoding/decoding process. Because the encoding and decoding drives cause an additional phase due to the Stark shift, we have to calibrate this phase. This is achieved by preparing qubit $Q_2$ in $\left(\ket g + \ket f\right)/\sqrt{2}$ state, then performing the encoding and decoding processes sandwiched with variable waiting times, and finally measuring the state tomography of qubit $Q_2$. We then can counteract this Stark shift by choosing a proper rotation axis of the decoding pulse in the process tomography measurement.

The two-photon transition drive of qubit $Q_1$ is calibrated in a similar way. The holonomic gates for Fock states $\ket 0$ and $\ket 1$ are realized when both the two-photon transition and the cavity-assisted Raman transition drives are simultaneously on, therefore the resonant frequencies of the two drives change slightly. We optimize the resonant frequencies of the two drives in order to realize high-fidelity holonomic gates.

\begin{figure}[tb]
\includegraphics{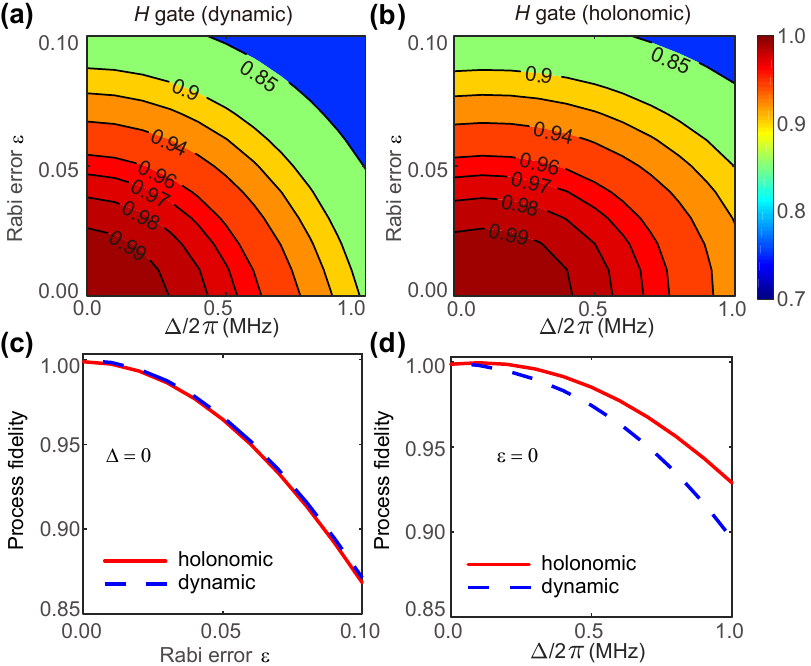}
\caption{Simulated performances for (a) dynamic and (b) holonomic Hadamard gate respectively as a function of Rabi error $\varepsilon$ (a relative offset in Rabi frequency) of the control field and qubit frequency detuning $\Delta$. (c) Process fidelity as a function of Rabi error $\varepsilon$ with $\Delta=0$. (d) Process fidelity as a function of qubit frequency detuning $\Delta$ with $\varepsilon=0$.}
\label{fig:simulation_H}
\end{figure}

\begin{figure}
\includegraphics{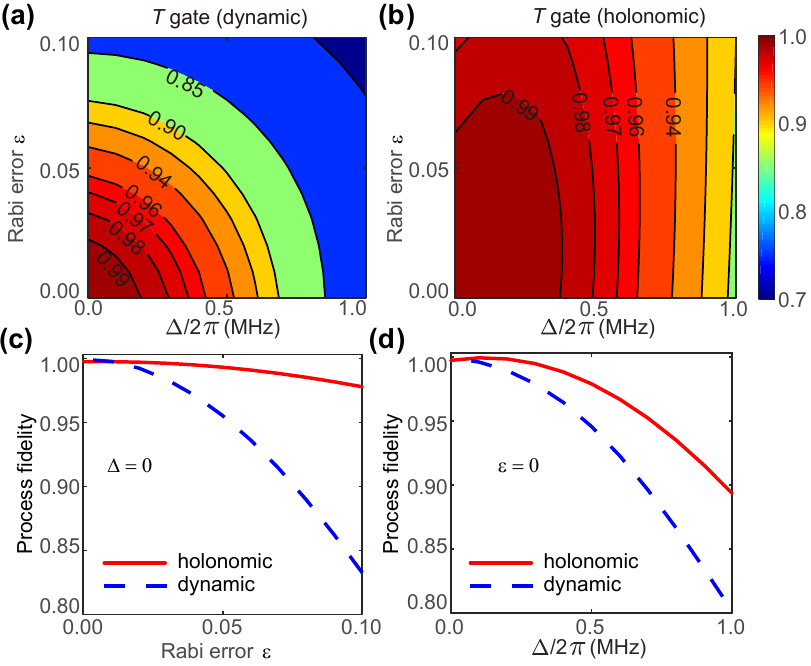}
\caption{Simulated performances for (a) dynamic and (b) holonomic $T$ gate respectively as a function of Rabi error $\varepsilon$ (a relative offset in Rabi frequency) of the control field and qubit frequency detuning $\Delta$. (c) Process fidelity as a function of Rabi error $\varepsilon$ with $\Delta=0$. (d) Process fidelity as a function of qubit frequency detuning $\Delta$ with $\varepsilon=0$.}
\label{fig:simulation_T}
\end{figure}

For the holonomic gates on $Q_1$, $X_{\pi}=U_2(\pi/2,0)$ and $Y_{\pi}=U_2(\pi/2,\pi/2)$ with $\theta=\pi/2$, we choose the effective coupling strengths $\tilde{g}_1/2\pi=\tilde{g}_2/2\pi=0.25$~MHz for both drives. Two square pulses (1420~ns long) with gradual ramp-up and ramp-down edges (10~ns for each) are used to implement these two holonomic gates. The only difference between these two gates is the phase difference between the two drives. For the two Hadamard gates $H_1=U_2(\pi/4,0)$ and $H_2=U_2(\pi/4,\pi/2)$ with $\theta=\pi/4$, the effective coupling strengths are $\tilde{g}_1/2\pi=0.25$~MHz and $\tilde{g}_2/2\pi=0.60$~MHz for the two-photon transition drive and the cavity-assisted Raman transition drive respectively. The time for these two Hadamard gates are 779~ns, also including the gradual ramp-up and ramp-down edges. Since the phase difference between the two drives can be well controlled, we could realize the holonomic gates $U_2(\theta,\phi)$ with arbitrary $\theta$ and $\phi$. With the same method for the transmon qubit, we can divide the evolution time into two intervals with different phase sets to realize arbitrary single-qubit holonomic gates $U_1(\theta,\gamma,\phi)$ on the Fock states.

\section{Noise-resilient feature of the holonomic gates}

Since the robustness of geometric phase gates against local noises has been extensively studied~\cite{zhu2005geometric,solinas2012on,johansson2012robustness} and experimentally verified~\cite{Berger2013Exploring,Yale2016Optical}, here we only consider the crosstalk noises. As qubit coherence properties are improved and the size of quantum system becomes larger, control amplitude errors and qubit frequency shifts induced by crosstalk (between qubits and their neighboring control drives; between neighboring qubits) become prominent gate error sources. To illustrate the robustness of holonomic gates against these errors, we numerically simulate the performance of dynamic and holonomic gates for the transmon qubit as a function of Rabi error $\varepsilon$ (a relative offset in Rabi frequency) of the control field and qubit frequency detuning $\Delta$. 

To realize the dynamic gates between $\ket{g}$ and $\ket{f}$, multiple sequential gate pulses are necessary. For example, for the single-qubit universal gate set, a Hadamard gate requires $X_{\pi}^{\mathrm{ge}}$$X_{\pi/2}^{\mathrm{ef}}$$X_{\pi}^{\mathrm{ge}}$, while a $T$ gate requires $X_{\pi}^{\mathrm{ge}}$$Y_{\pi}^{\mathrm{ef}}$$R_{\pi}^{\mathrm{ef}}$$X_{\pi}^{\mathrm{ge}}$, where $R_{\pi}^{\mathrm{ef}}$ denotes a $\pi$ rotation between $\ket{e}\leftrightarrow\ket{f}$ along an axis in the $xy$ plane with an angle of $-\pi/8$ to the $x$ axis. In contrast, the holonomic gate requires multiple drives to be applied simultaneously in a single step. This difference already makes holonomic gate more robust against qubit decoherence during gate operations. It is worth mentioning that for the dynamic gates extra phases caused by the energy space difference ($\omega_{\mathrm{ge}}\neq\omega_{\mathrm{ef}}$) need to be carefully calibrated. In our numerical simulation, we focus only on the crosstalk noise, so both $T_1$ and $T_2$ are set to be infinite. The pulses in the dynamic gates are set to have the same Gaussian envelope ($\sigma=30$~ns) with DRAG as those in the holonomic gates.  In Fig.~\ref{fig:simulation_H} (Fig.~\ref{fig:simulation_T}), we present the comparison between holonomic and dynamic Hadamard ($T$) gate, confirming the holonomic gates indeed perform better than the dynamic gates. This comparison can be generalized to the holonomic gates on the cavity, which provide an alternative way of arbitrary control over Fock states and could also be robust against experimental noises.

%As mentioned in the main text, the \{$\ket{g}, \ket{f}$\} encoding is important for distributed quantum information processing~\cite{pechal2014microwave,Kurpiers2017Deterministic}.

%\bibliography{bibliography}